\documentclass[aps,onecolumn,showpacs]{revtex4}
\usepackage{graphicx}
\usepackage[all]{xy}
\usepackage{amsmath}
\usepackage{amssymb}
\usepackage{epstopdf}
\usepackage[utf8]{inputenc}
\usepackage[T1]{fontenc} 
\newcommand{\be}{\begin{equation}}
\newcommand{\ee}{\end{equation}}
\newcommand{\bn}{\begin{eqnarray}}
\newcommand{\en}{\end{eqnarray}}
\newcommand{\bes}{\begin{subequations}}
\newcommand{\ees}{\end{subequations}}

\newcommand{\bb}{\bibitem}

\begin{document}

\title{The Extension Method for Bloch Branes}
\author{F. A. Brito$^{1\,2}${\footnote{email: fabrito@df.ufcg.edu.br }}, L. Losano$^{2}${\footnote{email: losano@fisica.ufpb.br }} and J. R. L. Santos$^{1}${\footnote{email: joaorafael@df.ufcg.edu.br}} }

\affiliation{{\small {
$^1$Unidade Acad\^emica de F\'isica, Universidade Federal de Campina Grande, 58429-900 Campina Grande, Para\'iba, Brazil \\
$^2$Departamento de F\'isica, Universidade Federal da Para\'iba, 58051-900 Jo\~ao Pessoa, Para\'iba, Brazil\\
}
}}

\begin{abstract}

The nature of gravity and its strength compared with other forces are fundamental challenges faced by the actual science. A path to understanding the hierarchy problem of gravity consists in adding an extra dimension to Einstein-Hilbert's general relativity. Therefore, gravity would be the only force leaking through this region of spacetime also known as brane. Despite the success of the mathematical description for the hierarchy issue through simple thin branes, it is also possible to derive a consistent theory of gravity using thick branes. The present work unveils a new procedure to determine analytic hybrid thick braneworld models. The discussions were based on the so-called Bloch branes, which are constructed via a two-field model coupled with gravity in 5-dimensional spacetime. Both fields used to build the Bloch branes depend only on the extra dimension, and the hybrid branes present internal structure, characterizing them as thick. In this work, we were able to derive new analytic thick brane models by using the extension method. Some of these new branes present a Minkowski spacetime immersed inside of the fifth-dimension. We showed how the symmetry of the emergent branes can be controlled via specific parameters. We also analyzed the stability conditions for each braneworld family, showing possible stable states for gravitons. 
\end{abstract}


\maketitle

\section{Introduction}
\label{sec:intro}

Over the last decades, the hierarchy problem has been at the center of the discussions on extensions of the standard model. A possible route addressed by Arkani-Hamed and Schmaltz \cite{arkani_99}, consists of using a thick brane to confine the Standard Model Fields. This approach allowed one to understand the fermion mass hierarchy and the proton stability as related with higher dimensional physics, instead due symmetries in high energy theories.  
Besides, an alternative path to understanding the hierarchy issue was introduced by Gogberashvili \cite{Gogberashvili/98, Gogberashvili/99}, and in the seminal papers of Randall and Sundrum \cite{rs_991, rs_992}, who used a new set of metric with an extra dimension to describe a brane. This metric has the property of being non-factorizable, there the standard four-dimensional metric is multiplied by a warp factor which grows exponentially as a function of the compactification radius. The so-called Randall-Sundrum braneworld model gave rise to a thin brane profile that contains the Standard Model fields, and it is an effective mechanism to reduce the energy scales from Planckian to TeV.

Several different proposals for generating models with thick branes have been presented in the literature so far \cite{thick_01, thick_02, thick_03, thick_04, Bernardini/2014, thick_051, thick_052, Gabriel/2014, Dutra/2015, thick_07, thick_08}, and most of them include the coupling of gravity with a scalar field, where the last depends only on the extra-dimension. One of these approaches is the so-called Bloch branes, introduced by Bazeia and Gomes \cite{bg} where the authors inspired themselves in a condensed matter model denominated Bloch walls (these are domain walls with internal structure), in the context of braneworlds scenarios. These branes are hybrid ones, which means that they are generated by multi-scalar field models (in this particular approach, a two scalar field model). There the authors were able to find analytic warp factor, as well as to verify the localization of gravity, by computing the graviton zero-mode. One feature related to the branes of such a model is that they are symmetric, and as it was pointed by \cite{dutra_13,bmm} a certain degree of asymmetry for the brane is desired to have models which are useful to deal with the hierarchy problem. In this work, we propose to generalize the initial Bloch branes approach by generating new exactly solvable braneworld models via an extension method introduced in \cite{bls}.

In our investigation, we are going to use some parameters to control the asymmetry of the brane. Moreover, we are going to verify carefully the stability of such branes, besides we also compute the zero-mode states for each new braneworld family. The set of ideas behind our investigation is summarized in the following nutshell: in section \ref{sec:gen} we describe some generalities about hybrid branes models. After that, in section \ref{sec:extmeth} we show the main ingredients necessary to implement the extension method. Moving to section \ref{sec:appl}, we apply the extension method in the braneworld scenario, and we exemplify it with three cases called $\phi^4$ versus $\chi^6$, $p$ versus $p$, and $\phi^4$ versus $p$. Section \ref{sec:stability} is dedicated to verifying the stability of each braneworld model derived in the previous section. Finally, we left our perspectives and final remarks for section \ref{sec:discuss}.

\section{Generalities}
\label{sec:gen}
In this section, we present some generalities concerning the application of scalar fields in braneworld scenarios, by following the recipe adopted in \cite{bg}.
Here, we studied the coupling between a two scalar field Lagrangian with gravity in $4+1$ dimensions, which means that in this representation the spacetime has an extra spatial coordinate named $y$. So, the action related to this previous Lagrangian is such that
\be \label{sec1_eq1}
S=\int\,d^4\,x\,dy\,\sqrt{|g|}\,\left[-\frac{R}{4}+{\cal L}(\phi,\partial_a{\phi};\chi,\partial_a\,\chi)\right]\,,
\ee
where we consider the standard form for ${\cal L}$, which means
\be \label{sec1_eq2}
{\cal L}=\frac{1}{2}\,\partial_a\,\phi\,\partial^{\,a}\,\phi+\frac{1}{2}\,\partial_a\,\chi\,\partial^{\,a}\,\chi-V(\phi\,\chi)\,.
\ee
In this approach we are working with $4\,\pi\,G=1$, $g$ is the determinant of the metric tensor $g_{\,ab}$, and the square of the line element is
\be \label{sec1_eq3}
ds_5^{\,2}=g_{a\,b}\,dx^{\,a}\,dy^{\,b}=e^{\,2\,A(y)}\,\eta_{\,\mu\,\nu}\,dx^{\,\mu}\,dx^{\,\nu}-dy^{\,2}\,,
\ee
with $a,b=0,1,...,4$, $\,\nu,\mu=0,1,...,3$, $\eta_{\,\mu\,\nu}=(1,-1,-1,-1)$. Besides $e^{\,2A}$ is the so-called warp-factor.  In order to guaranty 4D gravity localization along the extra dimension, the warp-factor needs to be integrable, which means that
\be \label{sec1_eq3_01}
\lim_{y\rightarrow \pm \infty}\,e^{\,2\,A(y)}=0\,.
\ee
Moreover, from Eq. \eqref{sec1_eq3} we yield to the 4D Minkowski spacetime by taking $y=0$, and for the asymptotic values of $y$, we have $ds^{\,2}=-dy^{\,2}$, which means a space-like interval in the fifth dimension.  The Einstein's equations for such a configuration are given by
\be \label{sec1_eq4}
G_{a\,b}=2\,T_{a\,b}\,,
\ee
where $T_{a\,b}$ is the energy momentum tensor in  $4+1$ dimensions. If we deal with  $\phi=\phi(y)$, $\chi=\chi(y)$, and $A=A(y)$, we can derive the following equations of motion for a braneworld model
\be \label{sec1_eq5}
\phi^{\,\prime\,\prime}+4\,A^{\,\prime}\,\phi^{\,\prime}=V_{\,\phi}\,; \qquad \chi^{\,\prime\,\prime}+4\,A^{\,\prime}\,\chi^{\,\prime}=V_{\,\chi}\,,
\ee
besides, once we are working with flat branes, the parameter $A$ must satisfy the equations
\be \label{sec1_eq6}
A^{\,\prime\,\prime}=-\frac{2}{3}\,\left(\phi^{\,\prime\,2}+\chi^{\,\prime\,2}\right)\,,
\ee
and
\be \label{sec1_eq7}
A^{\,\prime\,2}=\frac{1}{6}\,\left(\phi^{\,\prime\,2}+\chi^{\,\prime\,2}\right)-\frac{1}{3}\,V\,,
\ee
where primes mean derivatives in respect to $y$. In order to implement the first-order formalism, let us establish the definitions 
\be \label{sec1_eq8}
A^{\,\prime}=-\frac{W}{3}\,; \qquad \phi^{\,\prime}=\frac{W_{\,\phi}}{2}\,; \qquad \chi^{\,\prime}=\frac{W_{\,\chi}}{2}\,,
\ee
where $W=W(\phi,\chi)$. Thus, the potential $V$ needs to obey the constraint
\be \label{sec1_eq9}
V(\phi,\chi)=\frac{1}{8}\,\left(W_{\,\phi}^{\,2}+W_{\,\chi}^{\,2}\right)-\frac{W^2}{3}\,,
\ee
furthermore, the energy density for this system is described by
\be \label{sec1_eq10}
\rho(y)=e^{\,2A(y)}\,\left[\frac{\phi^{\,\prime\,2}}{2}+\frac{\chi^{\,\prime\,2}}{2}+V(\phi,\chi)\right]\,.
\ee

So, once we solve the first-order differential equations for $\phi$ and $\chi$, we can determine an analytic warp-factor. However, the main difficult concerning analytic results for two field models is that in general, these first-order differential equations are coupled, which make them hard to be integrated. 

To investigate the asymptotic structure of the spacetime where the branes are embedded, we find the general form of the Ricci scalar of our braneworlds metric as follows:
\be \label{ricci-scalar0}
R=-20 A'^2 - 8 A'',
\ee
that can still be written in terms of the scalar fields and potential according to Eqs.~(\ref{sec1_eq6})-(\ref{sec1_eq7}), i.e., 
\be \label{ricci-scalar}
R=2\phi'^2+2\chi'^2+\frac{20}3V.
\ee
At asymptotic behavior $\phi\to const.$ and $\chi\to const.$, and we achieve the scalar curvature $R$ totally governed by the behavior of the scalar potential  $V(\phi,\chi)$. This is precisely in agreement  with the formula $R=2n/(n-2)\Lambda$, for maximally symmetric spaces, with $V=\Lambda/2$, and $n=5$ spacetime dimensions.
We shall apply this result in the examples below, by just investigating the asymptotic behavior the scalar potential.

\section{Extension Method}
\label{sec:extmeth}

With the purpose to determine new analytic effective two scalar fields models in braneworld scenarios, we are going to use the extension method presented by \cite{bls}. In order to apply such a method, let us verify that the last two equations of \eqref{sec1_eq8} can be rearranged as
\be \label{sec2_eq1}
\phi_{\,\chi}=\frac{\phi^{\,\prime}}{\chi^{\,\prime}}=\frac{W_{\,\phi}(\phi,\chi)}{W_{\,\chi}(\phi,\chi)}\,,
\ee
whose integration lead us to analytic orbits relating fields $\phi$, and $\chi$. Before we discuss the generalities about the extension procedure, we may describe some main points related to the so-called deformation method introduced by \cite{blm}. 

Given two standard classical field theory Lagrangians
\be \label{sec2_eq2}
{\cal L}=\frac{1}{2}\,\partial_{\,\mu}\,\phi\,\partial^{\,\mu}\,\phi-V(\phi)\,; \qquad {\cal L}_d =\frac{1}{2}\,\partial_{\,\mu}\,\chi\,\partial^{\,\mu}\,\chi-\widetilde{V}(\chi)\,,
\ee
where the real fields $\phi$, and $\chi$ are static, one-dimensional, and need to obey the equations of motion
\be \label{sec2_eq3}
-\phi^{\,\prime\,\prime}+V_{\,\phi}=0\,; \qquad -\chi^{\,\prime\,\prime}+\widetilde{V}_{\,\chi}=0\,.
\ee
Then, by applying the BPS method, the solutions of these second-order differential equations may satisfy
\be \label{sec2_eq4}
\phi^{\,\prime}=W_{\,\phi}(\phi)\,; \qquad \chi^{\,\prime}=W_{\,\chi}(\chi)\,,
\ee
where we assumed that $V$ and $\widetilde{V}$ are defined as
\be \label{sec2_eq5}
V=\frac{W_{\,\phi}^{\,2}}{2}\,; \qquad \widetilde{V}=\frac{W_{\,\chi}^{\,2}}{2}\,.
\ee
These two different models may be connected via a specific function $f$, which is called deformation function. Such a mapping requires that  $\phi=f(\chi)$, which leads us to 
\be \label{sec2_eq601}
\phi^{\prime}=f_{\chi}\,\chi^{\prime}\,;\qquad W_{\phi}(\phi)=W_{\,\chi}(\chi)\,f_{\,\chi}\,,
\ee
meaning that
\be \label{sec2_eq6}
\frac{\phi^{\prime}}{\chi^{\prime}}=f_{\chi}\,; \qquad \phi_{\,\chi}=f_{\,\chi}=\frac{W_{\,\phi}}{W_{\,\chi}}\bigg|_{\phi\rightarrow\chi}\,.
\ee
We can observe that the last equation is very similar to \eqref{sec2_eq1}. Therefore, let us suppose that \eqref{sec2_eq6} can be rewritten as 
\be \label{sec2_eq7}
\phi_{\,\chi}=\frac{W_{\,\phi}}{W_{\,\chi}}=\frac{a_1\,W_{\,\phi}(\chi)+a_2\,W_{\,\phi}(\phi,\chi)+a_3\,W_{\,\phi}(\phi)+c_1\,g(\chi)+c_2\,g(\phi,\chi)+c_3\,g(\phi)}{b_1\,W_{\,\chi}(\chi)+b_2\,W_{\,\chi}(\phi,\chi)+b_3\,W_{\,\chi}(\phi)}\,,
\ee
where $W_{\phi}(\chi)$, $W_{\phi}(\phi,\chi)$, and $W_{\phi}(\phi)$ are equivalent since they are constructed with the deformation function $f(\chi)$, and with its inverse. In order to clarify this last sentence, let us consider , for instance, that
\begin{equation}
W_{\phi}(\phi) = \phi^{\,4}\,; \qquad \mbox{and} \qquad \phi=\chi^{\,1/2}\,,
\end{equation}
then
\begin{equation}
W_{\phi}(\chi)= \chi^{\,2}\,; \qquad W_{\phi} (\phi,\chi) = \phi^{\,2}\,\chi\,.
\end{equation}

The analogous procedure is taken for the different forms of $W_{\chi}$, and for the extra function $g$. Such a methodology corresponds to the simplest path to combine fields $\phi$, and $\chi$, moreover, it was successfully applied in cosmological scenarios \cite{ms_14,ms_18}, and in the description of crystalline polyethylene molecule \cite{smb}. 

This last function is used to connect fields $\phi,$ and $\chi$ in the effective two scalar field model. Moreover, the complete equivalence of the previous equation with  \eqref{sec2_eq6} imposes the constraints $a_1+a_2+a_3=1$, $b_1+b_2+b_3=1$, and $c_1+c_2+c_3=0$.

Consequently, once Eq. \eqref{sec2_eq7} has the same structure of Eq. \eqref{sec2_eq1}, we are allowed to identify 
\be \label{sec2_eq8}
W_{\,\phi}=a_1\,W_{\,\phi}(\chi)+a_2\,W_{\,\phi}(\phi,\chi)+a_3\,W_{\,\phi}(\phi)+c_1\,g(\chi)+c_2\,g(\phi,\chi)+c_3\,g(\phi)\,; 
\ee
\be \label{sec2_eq9}
W_{\,\chi}=b_1\,W_{\,\chi}(\chi)+b_2\,W_{\,\chi}(\phi,\chi)+b_3\,W_{\,\chi}(\phi)\,,
\ee
where $W_{\,\phi}$, and $W_{\,\chi}$ must satisfy the property
\be  \label{sec2_eq10}
W_{\,\chi\,\phi}=W_{\,\phi\,\chi}\,,
\ee
leading us to a constraint for the function $g$, given by
\be  \label{sec2_eq11}
b_2\,W_{\,\chi\,\phi}(\phi,\chi)+b_3\,W_{\,\chi\,\phi}(\phi)=a_1\,W_{\,\phi\,\chi}(\chi)+a_2\,W_{\,\phi\,\chi}(\phi,\chi)+c_1\,g_{\,\chi}(\chi)+c_2\,g_{\,\chi}(\phi,\chi)\,.
\ee
So, in order to have an unique form of $g$, we must choose either $c_1$ or $c_2$ equals to zero. Another interesting feature about this extension method is that the analytic solutions of the deformed one field systems are going to automatically satisfy the equations of motion related to the effective two scalar field model.  It is relevant to mention that the non-trivial contributions from the two scalar fields models for physical parameters, come from the interacting terms of the superpotential $W$ involving both fields.  From Eq. \eqref{sec2_eq7}, one can see that these terms are related to $a_2$ and $b_2$ constants. Besides, the mapping $\phi = f(\chi)$ is also known in the literature as orbit equation, and enables one to analytically solve the equations of motion of two scalar fields models \cite{Dutra/2005}. This mapping also allows one to reconstruct a single field model embedding the non-trivial contributions due the interacting terms, as one can see in the work of Chumbes et al. \cite{Chumbes/2009}.

\section{Application to Braneworld Scenarios}
\label{sec:appl}

\subsection{Example I - kink versus kink }
Let us firstly apply the extension method by considering the first-order differential equation
\be \label{sec3_eq1}
\phi^{\,\prime}=\frac{1}{2}\,W_{\,\phi}(\phi)=a^{\,2}-(\phi-a)^{\,2}\,,
\ee
whose analytic solution is
\be \label{sec3_eq2}
\phi(y)=a+a\,\tanh(a\,y)\,.
\ee
In classical field theory such a model describes a kink-like solution, corresponding to the $\phi^4$ potential \cite{Bazeia/2013}.
Now, if we work with the deformation function
\be \label{sec3_eq3}
\phi=f(\chi)=2\,a-\frac{a}{b^2}\,\chi^{\,2}\,,
\ee
we directly obtain the first-order differential equation
\be \label{sec3_eq4}
\chi^{\,\prime}=\frac{1}{2}\,W_{\,\chi}(\chi)=-\frac{a}{2}\,\chi\,\left(2-\frac{\chi^{\,2}}{b^{\,2}}\right)\,,
\ee
and the last one has 
\be \label{sec3_eq5}
\chi(y)=b\,\sqrt{1-\tanh(a\,y)}\,,
\ee
as its analytic solution. It  is relevant to point that in standard classical field theory this model is the kink solution of the $\chi^6$ potential \cite{Almeida/2004}.

Therefore, we are able to use $f$, and its inverse to write the relations
\be \label{sec3_eq6}
\frac{1}{2}\,W_{\,\phi}(\phi)=a^{\,2}-(\phi-a)^{\,2}\,; \qquad \frac{1}{2}\,W_{\,\phi}(\chi)=\frac{a^2}{b^2}\,\left(2\,\chi^{\,2}-\frac{\chi^{\,4}}{b^2}\right)\,; \qquad \frac{1}{2}\,W_{\phi}(\phi,\chi)=\frac{a^2}{b^2}\,\chi^{\,2}\,\left(\frac{\chi^{\,2}}{b^{\,2}}+2\,\frac{\phi-a}{a}\right)\;
\ee
\be \label{sec3_eq7}
\frac{1}{2}\,W_{\,\chi}(\chi)=-\frac{a}{2}\,\chi\,\left(2-\frac{\chi^{\,2}}{b^{\,2}}\right)\,;\qquad \frac{1}{2}\,W_{\,\chi}(\phi,\chi)=-\frac{a}{2}\,\chi\,\left(1+\frac{\phi-a}{a}\right)\,,
\ee
where we are not considering the form $W_{\,\chi}(\phi)$, since we would like to avoid terms involving rational powers in the final form of our potential. Such a consideration is equivalent to take $b_3=0$ in the extension method. Moreover, the constraint \eqref{sec2_eq11} with $c_1=0$ yields to
\be \label{sec3_eq8}
\frac{c_2}{2}\,g(\phi,\chi)=-\frac{b_2}{4}\,\chi^{\,2}-a_2\,\frac{a^2}{b^2}\,\chi^{\,2}\,\left(\frac{\chi^{\,2}}{b^{\,2}}+2\,\frac{\phi-a}{a}\right)-a_1\,\frac{a^{2}}{b^{2}}\,\left(2\,\chi^{\,2}-\frac{\chi^{\,4}}{b^{\,2}}\right)\,,
\ee
and we can use $\chi=f^{\,-1}(\phi)$ to rewrite this last equation as
\be \label{sec3_eq9}
\frac{c_2}{2}\,g(\phi)=-b_2\,\frac{b^2}{4}\,\left(1-\frac{\phi-a}{a}\right)-a^2(a_1+a_2)\,\left(1-\frac{(\phi-a)^{\,2}}{a^2}\right)\,.
\ee

\begin{figure}[h!]
\vspace{1cm}
\includegraphics[{height=04cm,angle=00}]{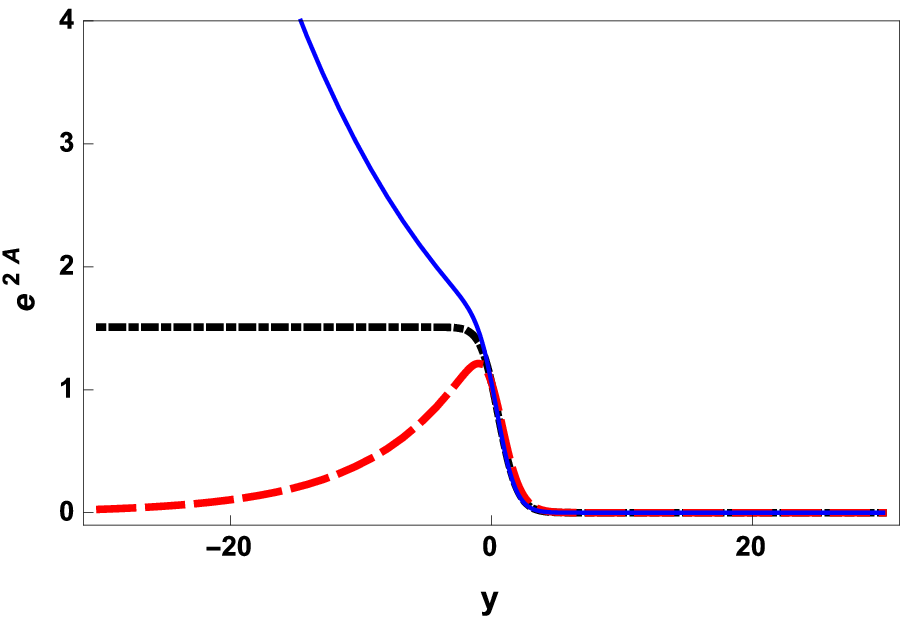} \hspace{0.1 cm} \includegraphics[{height=04cm,angle=00}]{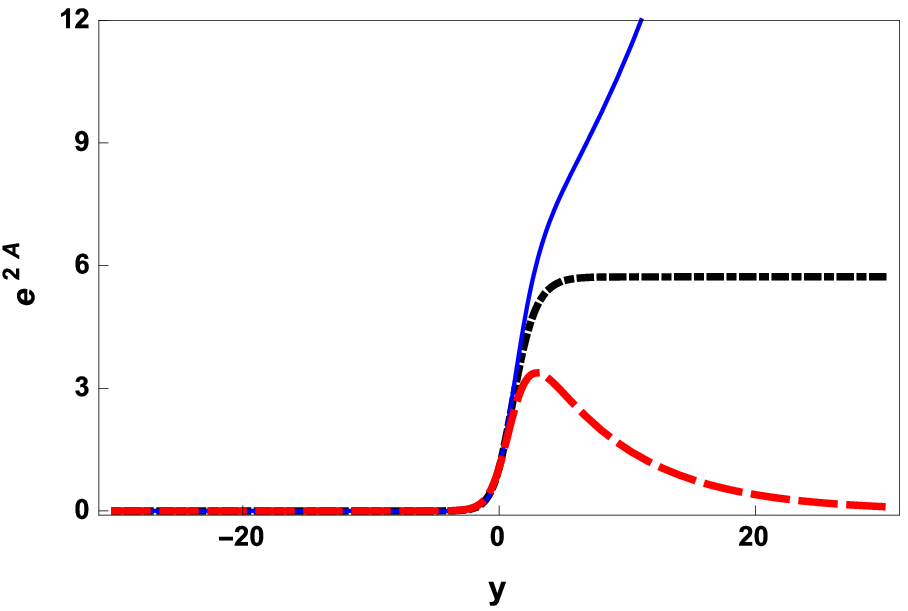}
\vspace{0.3cm}
\caption{In the left panel we see the warp-factor, for $a=1/2$, $b=2$, $b_2=1$ (dotted black curve), $b_2=0.9$ (dashed red curve), and $b_2=1.05$ (solid blue curve), corresponding to critical,  non-critical, and supercritical branes, respectively. In the right we worked with  $a=1/2$, $b=2$, $b_2=-1/6$ (dotted black curve), $b_2=-1/6+0.1$ (dashed red curve), and $b_2=-1/6-0.05$ (solid blue curve).}
\label{FIG1}
\end{figure}

\begin{figure}[h!]
\vspace{1cm}
\includegraphics[{height=04cm,angle=00}]{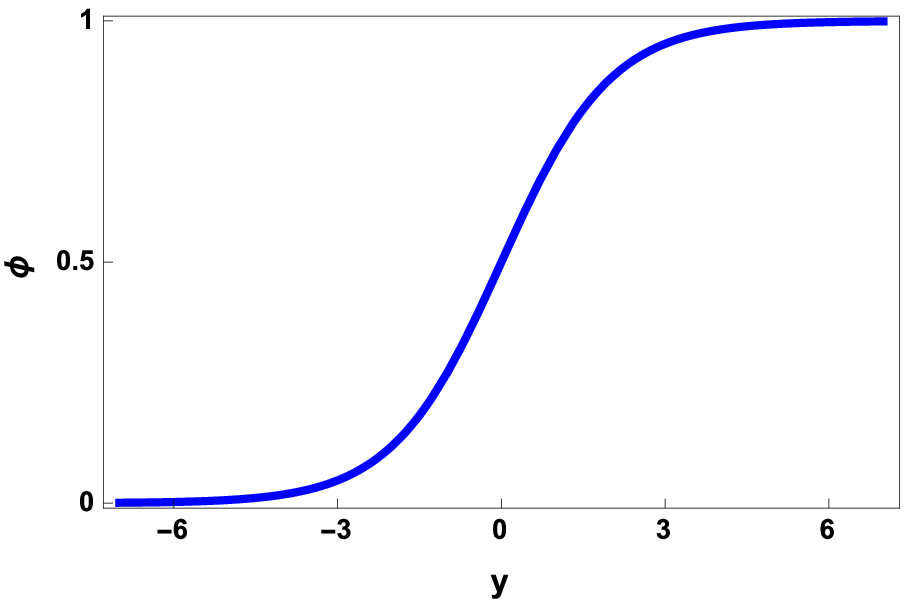} \hspace{0.1 cm} \includegraphics[{height=04cm,angle=00}]{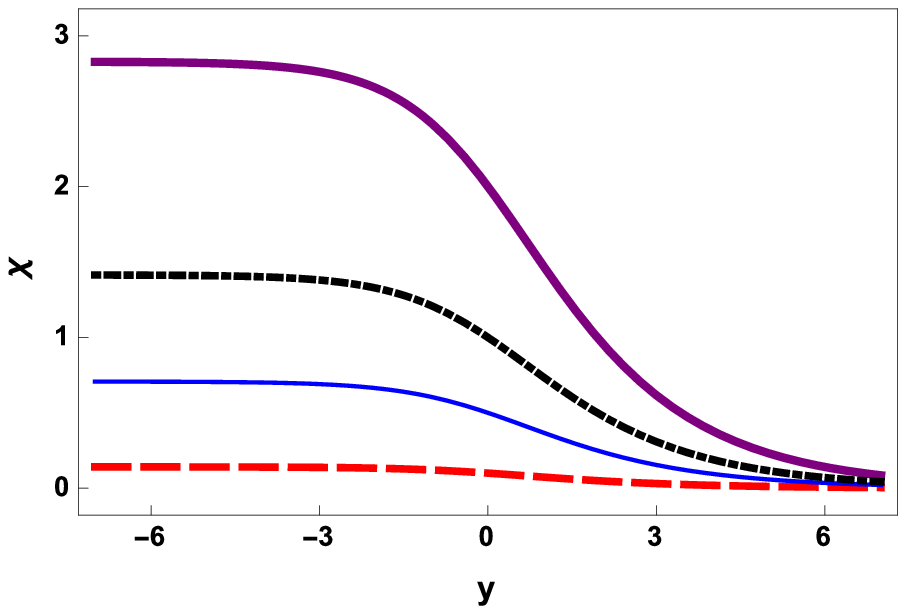}
\vspace{0.3cm}
\caption{The left panel shows $\phi$ for $a=1/2$. In the right frame we see $\chi$ for $a=1/2$, $b=2$ (thicker violet curve), $b=0.1$ (dashed red curve), $b=0.5$ (solid blue curve) and $b=1.0$ (dotted dashed black curve).}
\label{FIG2}
\end{figure}

Taking all these ingredients into \eqref{sec2_eq8}, and \eqref{sec2_eq9}, we find
\be \label{sec3_eq10}
W(\phi,\chi)=-(1-b_2)\,a\,\left(\chi^{\,2}-\frac{\chi^{\,4}}{4\,b^{\,2}}\right)-\frac{b_2}{2}\,\phi\,\chi^{\,2}+\frac{b_2\,b^2}{2}\,\left(2\phi-\frac{\phi^{\,2}}{2\,a}\right)+2\,\left(a\,\phi^{\,2}-\frac{\phi^{\,3}}{3}\right)\,,
\ee
and it is interesting to observe that if we choose $b=\pm\sqrt{2}$, $a=\pm1/2$, and $b_2=1$, the function $W$ has the form
\be \label{sec3_eq11}
\frac{W}{2}=-\frac{1}{4}\,\phi\,\chi^{\,2}+\phi-\frac{\phi^{\,3}}{3}\,,
\ee
which is the well known BNRT model for $r=1/4$ \cite{bnrt}. Moreover, as we can see in Eq. \eqref{sec3_eq10}, constant $b_2$ is multiplying the interacting term $\phi\,\chi^2$. Such a term is responsible for non-trivial contributions from the two scalar fields  in  the characterization of the branes. Therefore, taking $b_2 \neq 0$ yields to interesting behaviors of the braneworld parameters, as we are going to present next. Now, let us go back to Eq. \eqref{sec1_eq8}, where we can substitute  $W(\phi,\chi)$ together with the analytic solutions presented in \eqref{sec3_eq2}, and \eqref{sec3_eq5}, to determine the warp function 
\be \label{sec3_eq12}
A(y)=\frac{1}{36} \left[-2 a y \left(8 a^2+3 b^2 (2 b_2-1)\right)-2 \left(8 a^2+3 b^2\right) \log \,[\cosh (a y)]+4 a^2 \text{sech}^2(a y)+3 b^2 \tanh (a y)\right]\,,
\ee
leading us to the warp-factors plotted in Fig. \ref{FIG1}. Moreover, in Fig. \ref{FIG2} we depicted the explicit forms of fields $\phi$, and $\chi$. The panels of Fig. \ref{FIG1} unveil three different brane regimes, here called critical (dotted black curves), non-critical (dashed red curves), and supercritical (solid blue curves). 
The critical behavior of the brane can be determined by taking the following limits
\be \label{sec3_eq12_01}
\lim_{y\rightarrow\,\pm\,\infty}\frac{d}{d\,y}\,e^{\,2\,A(y)}=0\,.
\ee
So, if we consider $A(y)$ given by \eqref{sec3_eq12}, we find the constraints
\be
b_2=1\,;\qquad b_2=-\frac{8}{3}\,\frac{a^{\,2}}{b^{\,2}}\,,
\ee
where the last result unveil the critical value for the free parameter $b_2$. In Fig. \ref{FIG1} we also realize that $b_2$ controls the symmetry of the brane, and that the integrable branes can be found in the interval $-\frac{8}{3}\,\frac{a^{\,2}}{b^{\,2}}< b_2<1$. Out of such an interval, the brane enters in a supercritical regime, as one can observe in the thin blue curves of Fig. \ref{FIG1}. Moreover,  taking the asymptotic behavior of the fields back into the braneworld potential, we yield to the following cosmological constants 
\begin{equation}\label{cc_01}
V\left(\phi(+\infty),\chi(+\infty)\right)= \Lambda_{5\,+} = -\frac{1}{27} \left(8 \,a^3+3\, a\, b^2 \, b_2\right)^2\,,
\end{equation}
\begin{equation} \label{cc_02}
V\left(\phi(-\infty),\chi(-\infty)\right)= \Lambda_{5\,-} = \frac{b^4}{96}\, \left(3 \,b_2^2-32 \,a^2 (b_2-1)^2\right)\,.
\end{equation}
The previous constants together with the constraint over $b_2$ for integrable branes,  unveil that for positive asymptotic values of the extra dimension, we always have an $AdS_5$  bulk, since $\Lambda_{5\,+}$ is always negative. However, from \eqref{cc_02}, we verify that $\Lambda_{5\,-}$ can be negative ($AdS_5$ bulk), positive ($dS_5$ bulk) or null ($M_5$ bulk). The bulk curvature for negative asymptotic values of $y$ coordinate depends on the features of our scalar fields. Therefore, we verify that the existence of two scalar fields results in the complexity of the bulk curvature. A similar discussion about the cosmological constant in braneworld models can be found in  \cite{thick_07}.

\begin{figure}[h!]
\vspace{1cm}
\includegraphics[{height=04cm,angle=00}]{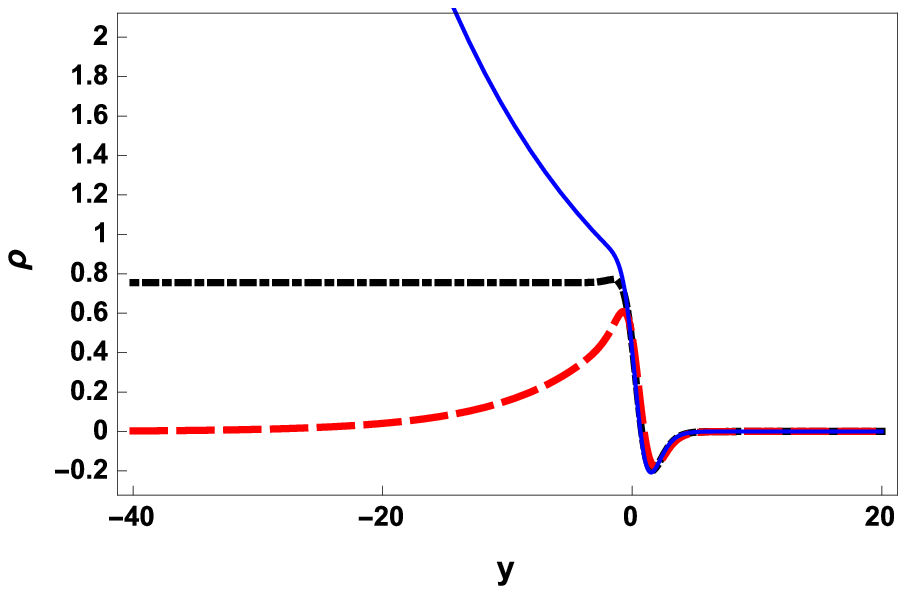} \hspace{0.1 cm} \includegraphics[{height=04cm,angle=00}]{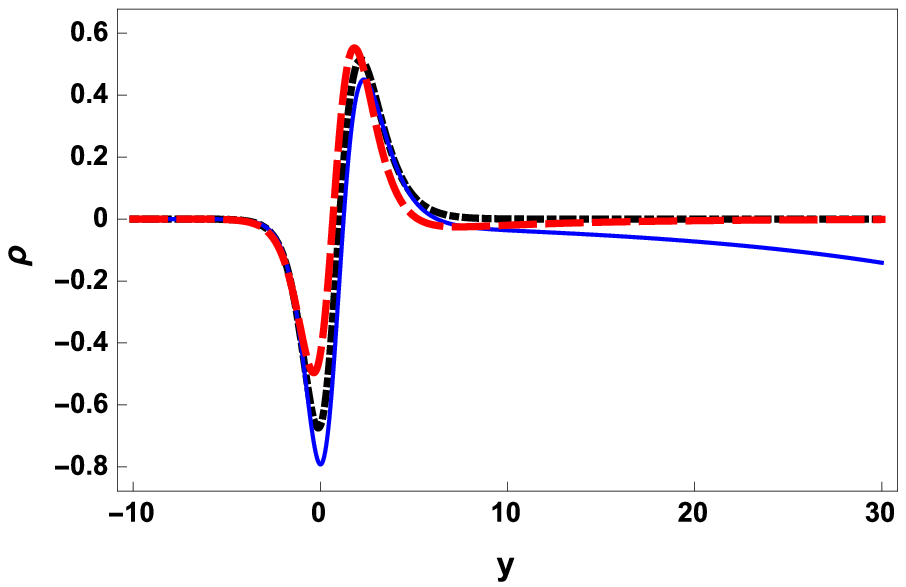}
\vspace{0.3cm}
\caption{In the left panel we depicted the energy density, for $a=1/2$, $b=2$, $b_2=1$ (dotted black curve), $b_2=0.9$ (dashed red curve), and $b_2=1.05$ (solid blue curve). In the right we choose  $a=1/2$, $b=2$, $b_2=-1/6$ (dotted black curve), $b_2=-1/6+0.1$ (dashed red curve), and $b_2=-1/6-0.05$ (solid blue curve).}
\label{FIG3}
\end{figure}

Withal, the energy density for such a model is
\bn \label{sec3_eq13} \nonumber
&&
\rho(y)=\bigg\{\frac{1}{2} a^4 \text{sech}^4(a y)-\frac{1}{432} a^2 \left(6 \left(4 a^2+b^2\right) \tanh (a y)-8 a^2 \tanh ^3(a y)+16 a^2+3 b^2 \tanh ^2(a y)+3 b^2 (4b_2-3)\right)^2 \\ \nonumber
&+&
\frac{1}{8} \left(\frac{1}{16} (\tanh (a y)-1)^2 \left(8 a^2 \tanh (a y)+8 a^2+b^2 b_2\right)^2+a^2 b^2 (\tanh (a y)+1) \text{sech}^2(a y)\right)+\frac{a^2 b^2 \text{sech}^4(a y)}{8 (1-\tanh (a y))}\bigg\} \\
&\times &
\exp\,\left[\frac{1}{18} \left(-2 a y \left(8 a^2+3 b^2 (2 \text{b2}-1)\right)-2 \left(8 a^2+3 b^2\right) \log (\cosh (a y))+4 a^2 \text{sech}^2(a y)+3 b^2 \tanh (a y)\right)\right]\,,
\en
which is illustrated in details in Fig. \ref{FIG3}. In such a figure we can verify that density $\rho$ is asymptotically null, excluding the critical case, and the supercritical cases. 


\subsection{Example II - $p$ model versus $p$ model}

In this example, we work with the first-order differential equation
\be \label{sec3_eq14}
\phi^{\,\prime}=\frac{W_{\,\phi}}{2}=p\,\left[\phi^{\,\frac{p-1}{p}}-\phi^{\,\frac{p+1}{p}}\right]\,,
\ee
whose analytic solution is
\be \label{sec3_eq15}
\phi(x)=\tanh^{\,p}(y)\,.
\ee
Then, by considering the simple deformation
\be \label{sec3_eq16}
\phi=\frac{\chi}{a}\,; \qquad \chi=a\,\tanh^{\,p}(y)\,,
\ee
we can rewrite \eqref{sec3_eq14} as 
\be
\chi^{\,\prime}=\frac{W_{\,\chi}}{2}=a\,p\,\left[\frac{\chi}{a}^{\,\frac{p-1}{p}}-\frac{\chi}{a}^{\,\frac{p+1}{p}}\right]\,,
\ee
which is satisfied by the second equation of \eqref{sec3_eq16}. So, repeating the methodology of the first example, we may rewrite $W_{\phi}$ and $W_{\chi}$ in the equivalent forms
\be
\frac{W_\phi(\phi)}{2}=p\,(\phi^{\frac{p-1}{p}}-\phi^{\frac{p+1}{p}}),\,\frac{W_\phi(\chi)}{2}=p\,\left[\left(\frac{\chi}{a}\right)^{\frac{p-1}{p}}-\left(\frac{\chi}{a}\right)^{\frac{p+1}{p}}\right],\,
\frac{W_\phi(\phi,\chi)}{2}=p\,\left[\left(\frac{\chi}{a}\right)^{\frac{p-1}{p}}-\phi\,\left(\frac{\chi}{a}\right)^{\frac{1}{p}}\right],
\ee
and
\be
\frac{W_\chi(\chi)}{2}=pa\left[\left(\frac{\chi}{a}\right)^{\frac{p-1}{p}}-\left(\frac{\chi}{a}\right)^{\frac{p+1}{p}}\right],\,
\frac{W_\chi(\phi)}{2}=pa(\phi^{\frac{p-1}{p}}-\phi^{\frac{p+1}{p}}),\,
\frac{W_\chi(\phi,\chi)}{2}=pa\left[\left(\frac{\chi}{a}\right)^{\frac{p-1}{p}}-\phi\left(\frac{\chi}{a}\right)^{\frac{1}{p}}\right].
\ee
In order to avoid negative exponent in the potential we choose $c_1=0$ and $b_3=0$, which means that $c_3=-c_2$, and $b_1+b_2=1$. Therefore, the function $g(\phi,\chi)$ for this case is
\be
\frac{c_2}{2}\,g(\phi,\chi)=-b_2\frac{\,p^2\,a^2}{p+1}\left(\frac{\chi}{a}\right)^{{\frac{p+1}{p}}} -a_2\,p\,\left[\left(\frac{\chi}{a}\right)^{{\frac{p-1}{p}}}-\phi\,\left(\frac{\chi}{a}\right)^{\frac{1}{p}}\right]-a_1\,p\,\left[\left(\frac{\chi}{a}\right)^{{\frac{p-1}{p}}}-\left(\frac{\chi}{a}\right)^{{\frac{p+1}{p}}}\right]\,.
\ee
By using the inverse of the deformation function, we can rewrite the last expression as
\be
\frac{c_2}{2}\,g(\phi)=-b_2\frac{\,p^2\,a^2}{p+1}\,\phi^{{\frac{p+1}{p}}}-(a_2+a_1)\,p\,(\phi^{\frac{p-1}{p}}-\phi^{\frac{p+1}{p}})\,. 
\ee

\begin{figure}[ht!]
\vspace{1cm}
\includegraphics[{height=04cm,angle=00}]{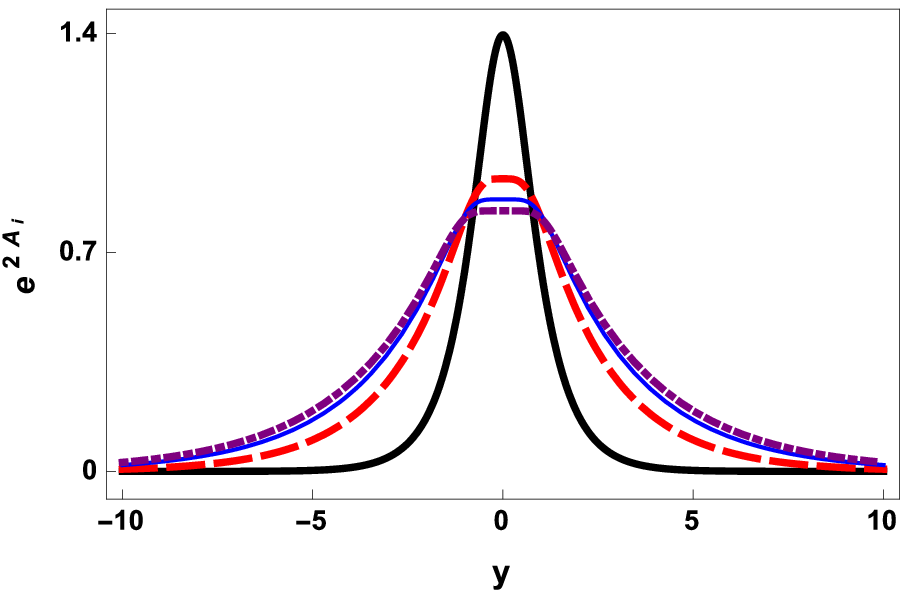} \hspace{0.1 cm} \includegraphics[{height=04cm,angle=00}]{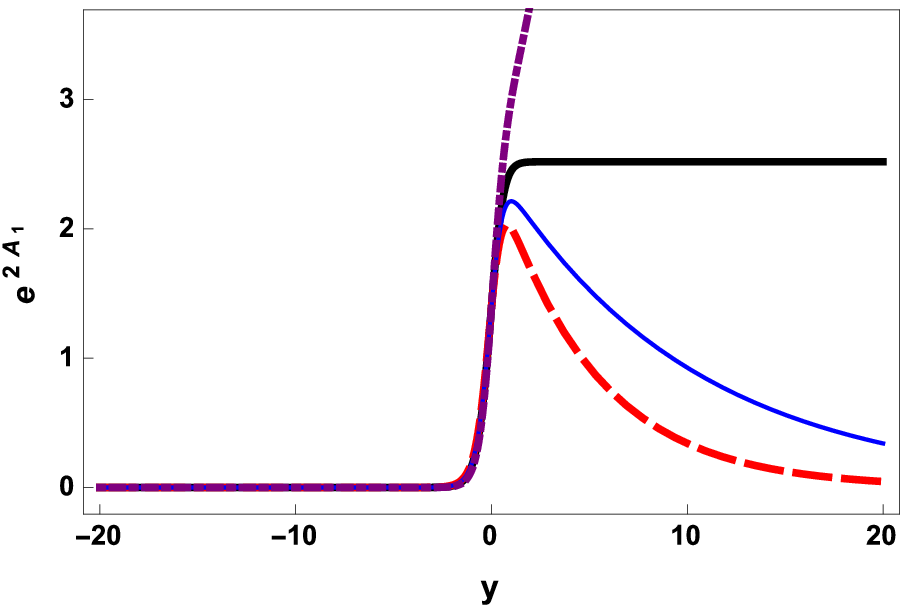}
\vspace{0.3cm}
\caption{In the left panel we see the warp-factors $e^{\,2\,A_i}$, with $i=1$ (solid black curve), $i=2$  (dashed red curve), $i=3$ (solid thin blue curve) and $i=4$ (dotted dashed violet curve), for $a=1$, $b_2=-0.5$, and $\widetilde{c}=0$. In the right we plotted $e^{\,2\,A_1}$ with $a=1$, $b_2=-0.5$, $\widetilde{c}=2/3$ (solid black curve), $\widetilde{c}=2/3-0.1$ (dashed red curve), $\widetilde{c}=2/3-0.05$ (solid thin blue curve), and $\widetilde{c}=2/3+0.1$ (dotted dashed violet curve).}
\label{FIG4}
\end{figure}

\begin{figure}[h!]
\vspace{1cm}
\includegraphics[{height=04cm,angle=00}]{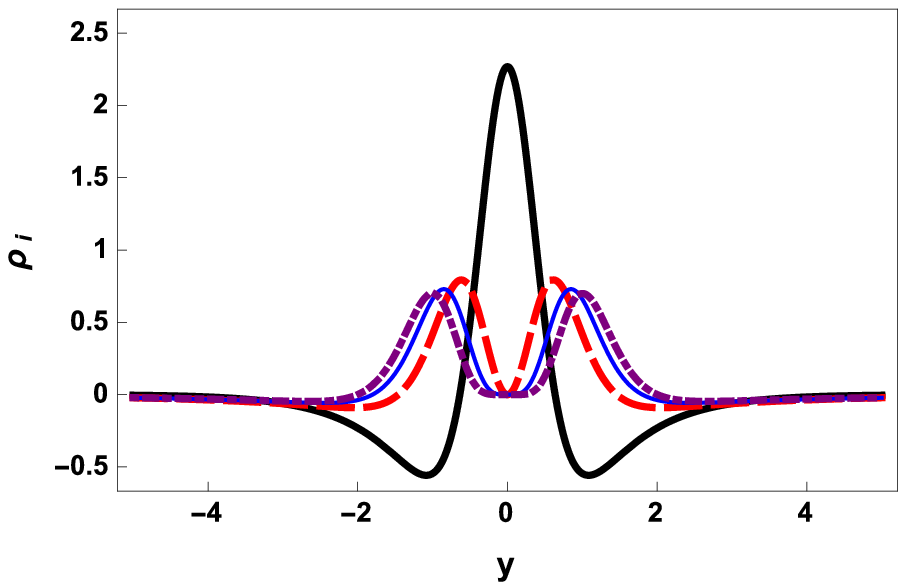} \hspace{0.1 cm} \includegraphics[{height=04cm,angle=00}]{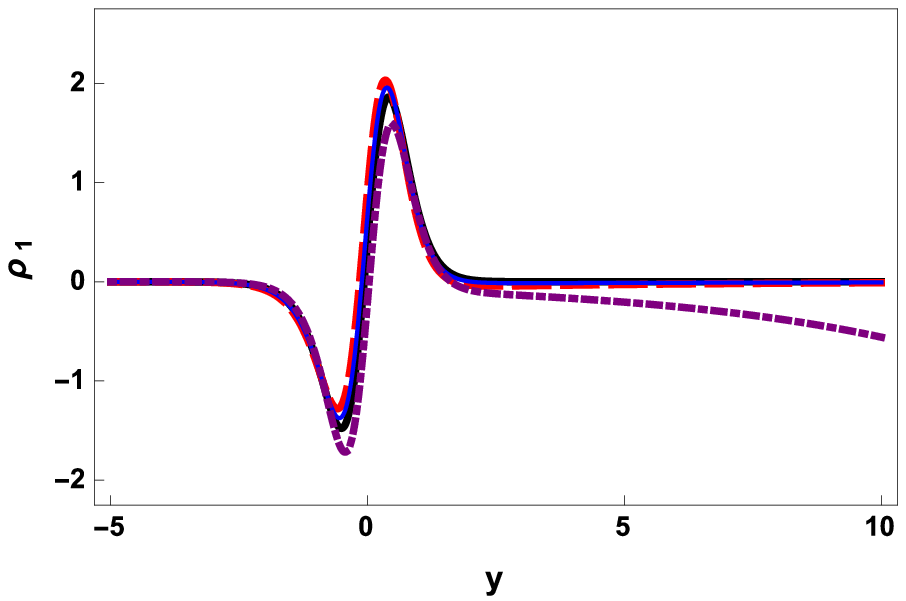}\hspace{0.1 cm} 
\vspace{0.3cm}
\caption{The left frame shows the energy density $\rho_i$, with $i=1$ (solid black curve), $i=2$  (dashed red curve), $i=3$ (solid thin blue curve) and $i=4$ (dotted dashed violet curve), for $a=1$, $b_2=-0.5$, and $\widetilde{c}=0$. In right frame we plotted $\rho_1$ with $a=1$, $b_2=-0.5$, $\widetilde{c}=2/3$ (solid black curve), $\widetilde{c}=2/3-0.1$ (dashed red curve), $\widetilde{c}=2/3-0.05$ (solid thin blue curve), and $\widetilde{c}=2/3+0.1$ (dotted dashed violet curve).}
\label{FIG5}
\end{figure}

Then, the application of the extension method yields to the effective superpotential:
\bn
W(\phi,\chi) &=& 2\,b_1\,p^2\,a^2\,\left[\frac{1}{2\,p-1}\,\left(\frac{\chi}{a}\right)^{\frac{2\,p-1}{p}}-\frac{1}{2\,p+1}\,\left(\frac{\chi}{a}\right)^{\frac{2\,p+1}{p}}\right]+2\,b_2\,p^3\,\frac{a^2\,\phi^{\frac{2\,p+1}{p}}}{(p+1)\,(2\,p+1)}\\ \nonumber
&&
+2\,b_2\,p^2\,a^2\,\left[\frac{1}{2\,p-1}\,\left(\frac{\chi}{a}\right)^{\frac{2\,p-1}{p}}-\frac{\phi}{p+1}\,\left(\frac{\chi}{a}\right)^{\frac{p+1}{p}}\right]+2\,p^2\,\left[\frac{\phi^{\frac{2\,p-1}{p}}}{2\,p-1}-\frac{\phi^{\frac{2\,p+1}{p}}}{2\,p+1}\right]-3\,\widetilde{c}\,,
\en
where $\widetilde{c}$ is an arbitrary real integration constant. Now, we can substitute this result together with the analytic solutions for $\phi$, and $\chi$ into \eqref{sec1_eq8} to determine that
\be
A^{\,\prime}=\frac{2 \left[p^2 \left(a^2 b_2+1\right)+1\right] \left[(2 p-1)\, \left(\tanh\,(y)\right)^{2 p+1}-(2 p+1)\,\left(\tanh\,(y)\right)^{2 p-1}\right]}{3 \left(4 p^2-1\right)}+\widetilde{c}\,.
\ee
Therefore, we present bellow the explicit forms of $A(y)$ for $p=1,2,3\,\mbox{and}\,4$, together with their respective energy densities:
\be
A_1=-\frac{1}{9} \left(a^2\,b_2+2\right) \left[4 \log (\cosh (y))-\text{sech}^2(y)\right]+\widetilde{c}\,y\,;
\ee
\be
A_2=-\frac{1}{90} \left(4 a^2\, b_2+5\right) \left[3 \text{sech}^4(y)-2 \text{sech}^2(y)+8 \log (\cosh (y))\right]+\widetilde{c}\,y\,;
\ee
\be
A_3=-\frac{1}{315} \left(9 a^2 \,b_2+10\right) \left[-5 \text{sech}^6(y)+12 \text{sech}^4(y)-3 \text{sech}^2(y)+12 \log (\cosh (y))\right]+\widetilde{c}\,y\,;
\ee
\be
A_4=-\frac{1}{18144}\left(16 a^2 \,b_2+17\right) \text{sech}^8(y) \left[11 \cosh (2 y)+72 \cosh (4 y)-3 \cosh (6 y)+384 \cosh ^8(y) \log (\cosh (y))+104\right]+\widetilde{c}\,y\,;
\ee
\bn
\rho_1 &=& \frac{1}{54} \left(\frac{27 \left(a^4 \left(b_2^2+1\right)+2 a^2 (b_2+1)+1\right) \text{sech}^4(y)}{a^2}-2 \left(2 \left(a^2\, b_2+2\right) \tanh (y) \left(\tanh ^2(y)-3\right)+9\, \widetilde{c}\right)^2\right) \\ \nonumber
&\times &
\exp \left(2\, \widetilde{c}\, y-\frac{2}{9} \left(a^2\, b_2+2\right) \left(4 \log (\cosh (y))-\text{sech}^2(y)\right)\right)\,;
\en
\bn \nonumber
\rho_2 &=&\left(\frac{\left(8 a^2 \left(2 a^2 \left(b_2^2+1\right)+b_2+4\right)+1\right) \tanh ^2(y) \text{sech}^4(y)}{8 a^2}-\frac{1}{675} \left(2 \left(4 a^2 b_2+5\right) \left(3 \tanh ^2(y)-5\right) \tanh ^3(y)+45 \,\widetilde{c}\right)^2\right) \\ 
&\times &
\exp \left(2\,\widetilde{c}\, y-\frac{1}{45} \left(4 a^2\,b_2+5\right) \left(3 \text{sech}^4(y)-2 \,\text{sech}^2(y)+8 \log (\cosh (y))\right)\right)\,;
\en
\bn \nonumber
\rho_3 &=&\left(\frac{\left(81 a^4 \left(b_2^2+1\right)+18 a^2 (b_2+9)+1\right) \tanh ^4(y) \text{sech}^4(y)}{18 a^2}-\frac{\left(2 \left(9 a^2 \,b_2+10\right) \left(5 \tanh ^2(y)-7\right) \tanh ^5(y)+105 \,\widetilde{c}\right)^2}{3675}\right) \\ 
&\times &
\exp \left(2 \, \widetilde{c}\, y-\frac{2}{315} \left(9 a^2\, b_2+10\right) \left(-5 \text{sech}^6(y)+12 \text{sech}^4(y)-3 \text{sech}^2(y)+12 \log (\cosh (y))\right)\right)\,;
\en
\bn
&& \nonumber
\hspace{-0.5cm}\rho_4=\left(\frac{\left(32 a^2 \left(8 a^2 \left(b_2^2+1\right)+b_2+16\right)+1\right) \tanh ^6(y) \text{sech}^4(y)}{32 a^2}-\frac{\left(2 \left(16 a^2 b_2+17\right) \left(7 \tanh ^2(y)-9\right) \tanh ^7(y)+189\, \widetilde{c}\right)^2}{11907}\right) \\ 
&&
\hspace{-0.5cm}\times\exp \left(2 \widetilde{c} y-\frac{\left(16 a^2 b_2+17\right) \text{sech}^8(y) \left(11 \cosh (2 y)+72 \cosh (4 y)-3 \cosh (6 y)+384 \cosh ^8(y) \log (\cosh (y))+104\right)}{9072}\right)
\en
whose details can be appreciated in Figs. \ref{FIG4} and \ref{FIG5}. There we see that when the integration constant $\widetilde{c}\neq 0$, we have an asymmetric warp-factor. Again, we are able to use Eq. \eqref{sec3_eq12_01} to find the critical branes for each one of our warp-factors. By taking $A_1$, we derive that the constraints for critical branes are
\be 
\widetilde{c}=\pm\,\frac{4}{9}\, \left(a^2\, b_2+2\right)\,.
\ee
So, we are able to compute integrable branes in the interval $ -\frac{4}{9}\, \left(a^2\, b_2+2\right) < \widetilde{c} < \frac{4}{9}\, \left(a^2\, b_2+2\right)$.

It is interesting to point that the warp-factor for $A_{4}$ (dotted dashed violet curve), depicted in the left panel of Fig. \ref{FIG4}, unveils a brane which is almost constant around $y=0$, so, in such a region the brane is close to the four-dimensional Minkowski spacetime, characterizing a bulk inside the fifth dimension. This behavior is corroborated by the energy density for this configuration (dotted dashed violet curve from Fig. \ref{FIG5} for $\widetilde{c} = 0$), which is almost null around $y=0$. 

Analogously to the first example, we can compute the cosmological constants using the asymptotic behavior of the scalar fields. Taking the braneworld potential for this case, we are able to find that 
\begin{equation}\label{cc_03}
V\left(\phi(+\infty),\chi(+\infty)\right)= \Lambda_{5\,+} = -\frac{\left(4 p^2 \left(a^2 b_2-3 \,\widetilde{c}+1\right)+3 \, \widetilde{c}+4\right)^2}{3 \left(1-4 p^2\right)^2}\,,
\end{equation}
\begin{equation} \label{cc_04}
V\left(\phi(-\infty),\chi(-\infty)\right)= \Lambda_{5\,-} = \frac{(a-1)^2}{2 a^4 p^2}-\frac{\left(4 p^2 \left(a^2 b_2+3 \,\widetilde{c}+1\right)-3\, \widetilde{c}+4\right)^2}{3 \left(1-4 p^2\right)^2}\,.
\end{equation}
These cosmological constants together with the constraint over $\widetilde{c}$ for integrable branes,  inform us that for positive asymptotic values of the extra dimension, we always have an $AdS_5$  bulk. Again, the curvature for the cosmological constant $\Lambda_{5\,-}$ depends on the features of our two scalar fields model, revealing the relevance of the scalar fields for controlling the bulk curvature. 

\subsection{Example III - kink versus $p$ model}

Another interesting analytic braneworld scenario raises from the coupling between the standard $\phi^4$ with a $p$ model. Here we consider the following kink-like profiles for the fields
\be
\phi=\tanh (y)\,; \qquad \chi= a\,\tanh^{\,p}(y)\,,
\ee
consequently,  the deformation function which connects these two models is
\be
\phi=\left(\frac{\chi}{a}\right)^{\,\frac{1}{p}}\,; \qquad \chi=a\,\phi^{\,p}\,.
\ee
So, we have the following equivalent representations for $W_{\phi}$ and $W_{\chi}$:
\be
\frac{W_{\phi}(\phi)}{2}=1-\phi^{\,2}, \qquad \frac{W_{\phi}(\chi)}{2}=1-\left(\frac{\chi}{a}\right)^{\,\frac{2}{p}},\qquad \frac{W_{\phi}(\phi,\chi)}{2}=\left(1-\left(\frac{\chi}{a}\right)^{\,\frac{1}{p}}\right)\,(1+\phi)\,; 
\ee
\be
\frac{W_\chi(\chi)}{2}=p\,a\,\left(\left(\frac{\chi}{a}\right)^{\,\frac{p-1}{p}}-\left(\frac{\chi}{a}\right)^{\,\frac{p+1}{p}}\right), \,\, \frac{W_{\chi}(\phi)}{2}=p\,a\,\left(\phi^{\,p-1}-\phi^{\,p+1}\right), \,\, \frac{W_{\chi}(\phi,\chi)}{2}=p\,a\,\left(\left(\frac{\chi}{a}\right)^{\,\frac{p-1}{p}}-\phi^{\,p}\,\left(\frac{\chi}{a}\right)^{\,\frac{1}{p}}\right)\,.
\ee
From the last expressions we are able to obtain
\be
\frac{c_2}{2}\,g(\phi,\chi)=b_3\,p\,a\,\left[(p-1)\,\phi^{p-2}-(p+1)\,\phi^{\,p}\right]\,\chi-\frac{b_2\,a^{\,2}\,p^{3}}{p+1}\phi^{\,p-1}\,\left(\frac{\chi}{a}\right)^{\frac{p+1}{p}}-a_1\,\left(1-\left(\frac{\chi}{a}\right)^{\,\frac{2}{p}}\right)-a_2\,\left(1-\left(\frac{\chi}{a}\right)^{\frac{1}{p}}\right)\,(1+\phi)\,
\ee
where we choose $c_1=0$. Moreover, using the inverse of the deformation function we directly determine
\be
\frac{c_2}{2}\,g(\phi)=b_3\,p\,a^{\,2}\,\left[(p-1)\,\phi^{\,p-2}-(p+1)\,\phi^{\,p}\right]\,\phi^{\,p}-\frac{b_2\,a^{\,2}\,p^{3}}{p+1}\,\phi^{\,2\,p}-(a_1+a_2)\,\left(1-\phi^{\,2}\right)\,.
\ee

\begin{figure}[ht!]
\vspace{1cm}
\includegraphics[{height=04cm,angle=00}]{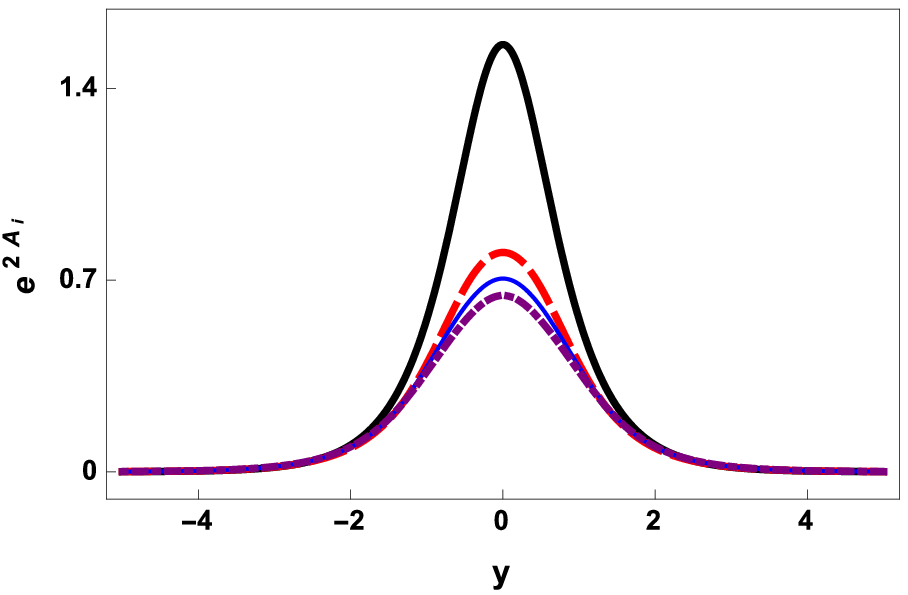} \hspace{0.1 cm} \includegraphics[{height=04cm,angle=00}]{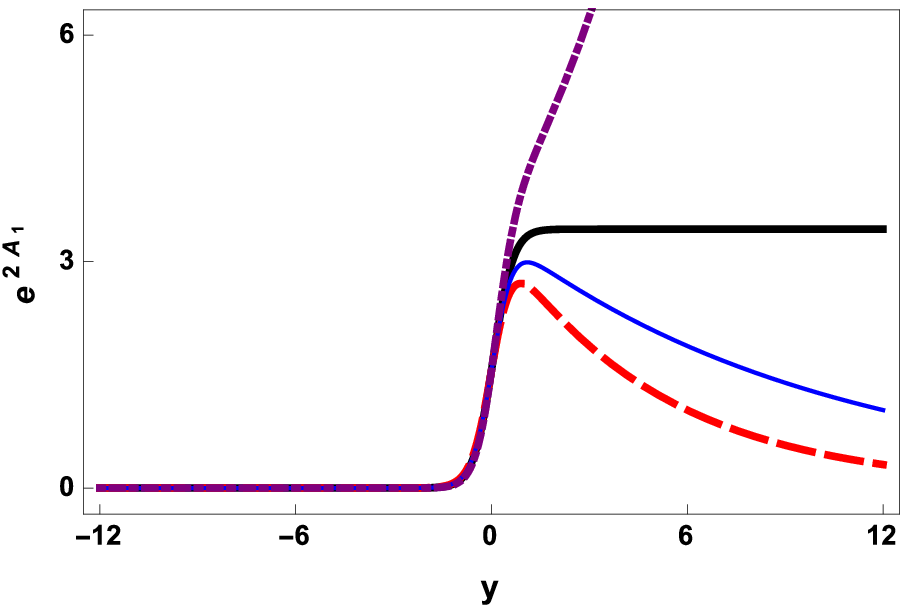}
\vspace{0.3cm}
\caption{In the left panel we see the warp-factors $e^{\,2\,A_i}$, with $i=1$ (solid black curve), $i=2$  (dashed red curve), $i=3$ (solid thin blue curve) and $i=4$ (dotted dashed violet curve), for $a=1$, and $\widetilde{c}=0$. In the right we plotted $e^{\,2\,A_1}$ with $a=1$, $\widetilde{c}=8/9$ (solid black curve), $\widetilde{c}=8/9-0.1$ (dashed red curve), $\widetilde{c}=8/9-0.05$ (solid thin blue curve), and $\widetilde{c}=8/9+0.1$ (dotted dashed violet curve).}
\label{FIG6}
\end{figure}

\begin{figure}[h!]
\vspace{1cm}
\includegraphics[{height=04cm,angle=00}]{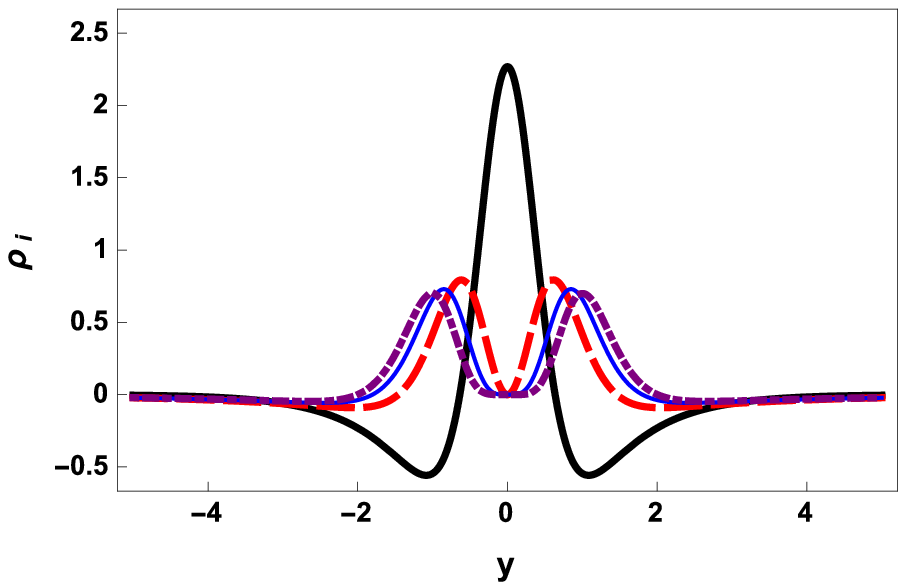} \hspace{0.1 cm} \includegraphics[{height=04cm,angle=00}]{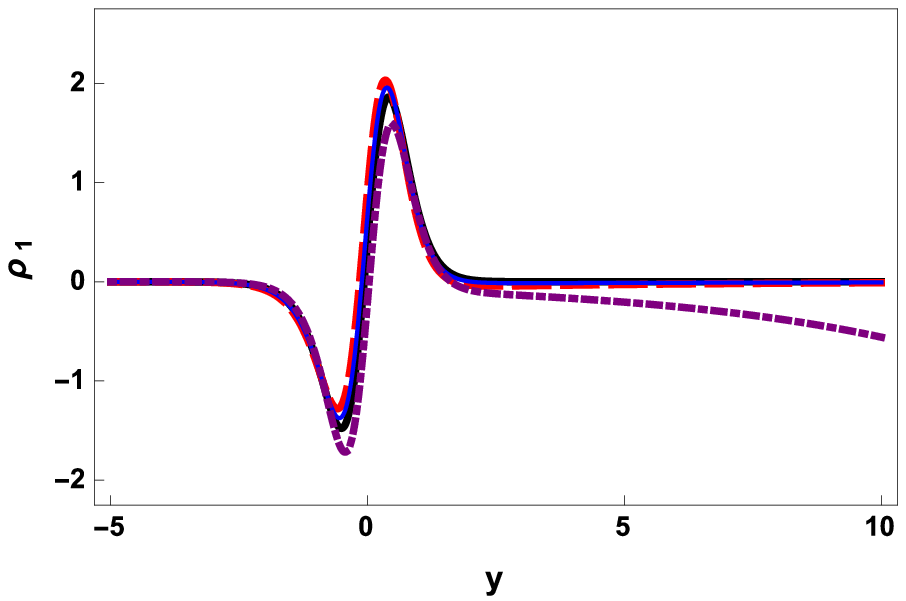}\hspace{0.1 cm} 
\vspace{0.3cm}
\caption{The left frame shows the energy density $\rho_i$, with $i=1$ (solid black curve), $i=2$  (dashed red curve), $i=3$ (solid thin blue curve) and $i=4$ (dotted dashed violet curve), for $a=1$, and $\widetilde{c}=0$. In right frame we plotted $\rho_1$ with $a=1$, $\widetilde{c}=8/9$ (solid black curve), $\widetilde{c}=8/9-0.1$ (dashed red curve), $\widetilde{c}=8/9-0.05$ (solid thin blue curve), and $\widetilde{c}=8/9+0.1$ (dotted dashed violet curve).}
\label{FIG7}
\end{figure}

These previous ingredients yield us to the effective superpotential
\bn
W(\phi,\chi)&=&2\,a^2 b_1 p^2 \left(\frac{\left(\frac{\chi }{a}\right)^{\frac{2 p-1}{p}}}{2 p-1}-\,\frac{\left(\frac{\chi }{a}\right)^{\frac{2 p+1}{p}}}{2 p+1}\right)+2\,\frac{a^2 b_2 p^3 \phi ^{2 p+1}}{(p+1) (2 p+1)}+2\,\frac{a^2 b_2 p^2 \left(\frac{\chi }{a}\right)^{\frac{2 p-1}{p}}}{2 p-1}-2\,\frac{a^2 b_2 p^2 \phi ^p \left(\frac{\chi }{a}\right)^{\frac{p+1}{p}}}{p+1} \\ \nonumber
&+&
2\,a^2 b_3 p \left(\frac{(p+1) \phi ^{2 p+1}}{2 p+1}-\,\frac{(p-1) \phi ^{2 p-1}}{2 p-1}\right)+2\,a\, b_3\, p \,\chi \, \left(\phi ^{p-1}-\phi ^{p+1}\right)-2\,\left(\frac{\phi ^3}{3}-\phi\right)-3\,\widetilde{c}\,,
\en
and by taking it back together with the static solutions to Eq. \eqref{sec1_eq8}, we find
\be
A^{\,\prime}= -\frac{2}{9} \text{sech}(y) \left(\frac{3 a^2 p^2 \text{csch}(y) (2 p+\cosh (2 y)) \tanh ^{2 p}(y)}{4 p^2-1}+(\cosh (2 y)+2) \tanh (y) \text{sech}(y)\right)+\widetilde{c}\,.
\ee
Therefore, the warp functions and their respective energy densities for $p=1,\,2,\,3\,\,\text{and}\,\,4$ are given by
\be
A_1=\frac{1}{9}\, \left(\left(a^2+1\right) \text{sech}^2(y)-4 \left(a^2+1\right) \log (\cosh (y))\right)+\widetilde{c}\, y\,;
\ee
\be
A_2=\frac{1}{45} \left(-6 a^2 \text{sech}^4(y)+\left(4 a^2+5\right) \text{sech}^2(y)-4 \left(4 a^2+5\right) \log (\cosh (y))-8 a^2\right)+\widetilde{c}\, y\,;
\ee
\be
A_3=\frac{1}{315} \left(\left(27 a^2+35\right) \text{sech}^2(y)-4 \left(27 a^2+35\right) \log (\cosh (y))-9 a^2 (6 \cosh (2 y)+1) \text{sech}^6(y)-54 a^2\right)+\widetilde{c}\, y\,;
\ee
\bn
A_4&=&\frac{1}{567}\, \bigg(-3\, \left(\left(64\, a^2+84\right)\, \log (\cosh (y))+32\, a^2\right)+\left(48 \,a^2+63\right)\, \text{sech}^2(y) \\ \nonumber
&&
-a^2\, (28\, \cosh (2 y)+45\, \cosh (4 y)+67)\, \text{sech}^8(y)\bigg)+\widetilde{c}\,y\,;
\en
\bn
\rho_1 &=&\frac{1}{27} (\cosh(y))^{-\frac{8}{9} \left(a^2+1\right)}\, e^{\frac{2}{9}\, \left(a^2+1\right)\, \text{sech}^2(y)+2\, \widetilde{c}\, y}\,\bigg(72 \left(a^2+1\right)\, \widetilde{c}\, \tanh (y)+36 \left(a^2+1\right)\, \widetilde{c}\, \tanh (y)\, \text{sech}^2(y) \\ \nonumber
&+&
4 \left(a^2+1\right)^2 \text{sech}^6(y)+3 \left(a^2+1\right) \left(4 a^2+13\right) \text{sech}^4(y)-16 \left(a^2+1\right)^2-81 \,\widetilde{c}^{\,2}\bigg)\,;
\en
\bn \nonumber
\rho_2 &=& \bigg(-\frac{1}{675} \left(24 a^2 \tanh ^5(y)+\left(10-40 a^2\right) \tanh ^3(y)+45 \,\widetilde{c}-30 \tanh (y)\right)^2-2 a^2 \text{sech}^6(y)+\text{sech}^4(y)\left(2 a^2 \tanh ^2(y)\right. \\ 
&+ &
\left. 2 a^2+1\right)\bigg)\exp \left(\frac{2}{45} \left(-6 a^2 \text{sech}^4(y)+\left(4 a^2+5\right) \text{sech}^2(y)-4 \left(4 a^2+5\right) \log (\cosh (y))-8 a^2+45\,\widetilde{c}\, y\right)\right)\,;
\en
\bn
\rho_3 &=& \left(\text{sech}^4(y) \left(9 a^2 \tanh ^4(y)+1\right)-\frac{\left(270 a^2 \tanh ^7(y)-378 a^2 \tanh ^5(y)+315\,\widetilde{c}+70 \tanh ^3(y)-210 \tanh (y)\right)^2}{33075}\right)  \\ \nonumber
&\times &
\exp \left(2 \left(\frac{1}{315} \left(\left(27 a^2+35\right) \text{sech}^2(y)-4 \left(27 a^2+35\right) \log (\cosh (y))-9 a^2 (6 \cosh (2 y)+1) \text{sech}^6(y)-54 a^2\right)+\widetilde{c}\, y\right)\right);
\en
\bn
\rho_4 &=&\frac{(\cosh(y))^{-\frac{8}{189} \left(16 a^2+21\right)}}{11907}\,\bigg((11907 \text{sech}^4(y) \left(16 a^2 \tanh ^6(y)+1\right)  \\ \nonumber
&-&
 \left(-32 a^2 (\cosh (2 y)+8) \tanh ^7(y) \text{sech}^2(y)+189\, \widetilde{c}+42 \tanh ^3(y)-126 \tanh (y)\right)^2\bigg) \\ \nonumber
& \times &
\exp \left(-\frac{2}{567} \left(-3 \left(16 a^2+21\right) \text{sech}^2(y)+a^2 (28 \cosh (2 y)+45 \cosh (4 y)+67) \text{sech}^8(y)+96 a^2-567\, \widetilde{c}\, y\right)\right)\,.
\en
The previous relations are shown in the panels of Fig. \ref{FIG6} and \ref{FIG7}, where we realize that asymmetric warp-factors appear if $\widetilde{c}$ is different of zero. Analogously to the previous examples, we can use the limits \eqref{sec3_eq12_01} to establish that the critical branes raise if
\be
\widetilde{c}=\pm\,\frac{4}{9}\,\left(a^2+1\right)\,,
\ee
which means that integrable branes are constrained in the interval $\frac{4}{9}\,\left(a^2+1\right) < \widetilde{c} < \frac{4}{9}\,\left(a^2+1\right)$.

By repeating the procedures adopted in the last two examples,  we can compute the cosmological constants for this scenario using the asymptotic behavior of the scalar fields. So, considering the braneworld potential for this case, we yield to
\begin{equation}\label{cc_05}
V\left(\phi(+\infty),\chi(+\infty)\right)= \Lambda_{5\,+} = -\frac{\left(4 p^2 \left(3 a^2-9\, \widetilde{c}+4\right)+9\,\widetilde{ c}-4\right)^2}{27 \left(1-4 p^2\right)^2}\,,
\end{equation}
\begin{eqnarray}\label{cc_06}
\nonumber
V\left(\phi(-\infty),\chi(-\infty)\right)= \Lambda_{5\,-} &=& \frac{1}{24} \Bigg(\frac{12 a^4 \left((-1)^p-1\right)^2 p^2 \left(b_2 \left((-1)^p+1\right) p^2 +b_3 (p+1) \left((-1)^p (p+1)+2\right)\right)^2}{(p+1)^2} \\ 
&&
-\frac{8 \left(-4 p^2 \left(3 a^2+9 \,\widetilde{c}+4\right)+9 \,\widetilde{c}+4\right)^2}{9 \left(1-4 p^2\right)^2}\Bigg)\,.
\end{eqnarray}
These cosmological constants together with the constraint over $\widetilde{c}$ for integrable branes show that for positive asymptotic values of the extra dimension, we always have an $AdS_5$  bulk. 
We also realize that for even values of $p$ the constant $\Lambda_{5\,-}$ is always negative (if we are considering integrable branes), which means that we have an $AdS_{5}$ bulk in this asymptotic regime.  If we work with odd values of $p$, then the curvature behavior will depend on the free parameters related to the scalar fields, and also on the extension method parameters $b_2$, and $b_3$.

\section{Stability of the Branes}
\label{sec:stability}

In order to discuss the stability of the gravitational sector for the braneworld scenarios, we are going to adopt the procedures used in ref. \cite{bg, almeida2014}. So, by considering a metric perturbation given by
\be \label{sec4_eq1}
ds^{\,2}=e^{\,2A(y)}\,\left(\eta_{\,\mu\,\nu}+\epsilon\,h_{\,\mu\,\nu}\right)\,dx^{\,\mu}\,dx^{\,\nu}-dy^2\,,
\ee
where $h_{\,\mu\,\nu}$ stands for the graviton with the axial gauge $h_{5\,N}=0$, and if {the metric fluctuation obeys the transverse traceless gauge} (which is denoted by  $\bar{h}_{\,\mu\,\nu}$), it is possible to prove that such a fluctuation needs to satisfy the differential equation \cite{bg, almeida2014, dewolfe/2000}
\be \label{sec4_eq2}
\bar{h}_{\,\mu\,\nu}^{\,\prime\,\prime}+4\,A^{\,\prime}\,\bar{h}_{\,\mu\,\nu}^{\,\prime}=e^{\,-2A}\,\square\bar{h}_{\,\mu\,\nu}\,.
\ee
Now, let us re-scale the coordinate $y$ by using the transformation $dz=e^{\,-A(y)}\,dy$, and let us also redefine $\bar{h}_{\,\mu\,\nu}$ as
\be \label{sec4_eq3}
\bar{h}_{\,\mu\,\nu}=e^{\,i\,k\,\cdot\,x}\,e^{\,-\frac{3}{2}\,A(z)}\,H_{\,\mu\,\nu}\,,
\ee
such that the substitution of the previous definitions in the last differential equation results in
\be \label{sec4_eq4}
-\frac{d^2\,H_{\,\mu\,\nu}}{dz^{\,2}}+U(z)\,H_{\,\mu\,\nu}=k^{\,2}\,H_{\,\mu\,\nu}\,.
\ee
Therefore, we have a Schroedinger-like equation with an effective potential
\be \label{sec4_eq5}
U(z)=\frac{3}{2}A_{\,z\,z}+\frac{9}{4}\,A_z^{\,2}\,.
\ee

The zero-mode ($k=0$) solution for \eqref{sec4_eq4} has the following general form
\be \label{sec4_eq6}
H_{\,\mu\,\nu}(z)=N_{\,\mu\,\nu}\,e^{\,\frac{3}{2}\,A(z)}\,,
\ee
where $N_{\,\mu\,\nu}$ is a normalization constant. We can use the prerogative established by \cite{campos2002},  to set the equality 
\be \label{sec4_eq7}
A_{z}=e^{\,A(y)}\,\frac{d\,A}{d\,y}\,; \qquad A_{zz}=\left(A^{\,\prime\,\prime}+A^{\,\prime\,2}\right)\,e^{\,2\,A(y)}\,,
\ee
which allows us to rewrite the Schroedinger-like potential as
\be \label{sec4_eq8}
U(y)=\frac{3}{4}\,e^{\,2\,A(y)}\,\left(2\,A^{\,\prime\,\prime}+5\,A^{\,\prime\,2}\right)\,.
\ee
Moreover, in order to write the zero modes in terms of $y$, we can apply the change of variable $\bar{h}_{\,\mu\,\nu}=e^{\,-2\,A(y)}\,\xi_{\,\mu\,\nu}(y)\,e^{\,i\,k\,\cdot\,x}$ in \eqref{sec4_eq2}, yielding to
\be \label{sec4_eq9}
\frac{d^{\,2}\xi_{\,\mu\,\nu}}{d\,y^{\,2}}-4\,A^{\,\prime\,2}\,\xi_{\,\mu\,\nu}-e^{\,-2\,A(y)}\,k^{\,2}\,\xi_{\,\mu\,\nu}=0\,,
\ee
then, the solution for the zero-mode has to obey the equation
\be \label{sec4_eq10}
\left(\frac{d}{d\,y}+2\,A^{\,\prime}\right)\,\left(\frac{d}{d\,y}-2\,A^{\,\prime}\right)\,\xi_{\,\mu\,\nu}=0\,.
\ee
Consequently, the zero-mode eigenstate is such that
\be \label{sec4_eq11}
\xi_{\,\mu\,\nu}=N_{\,\mu\,\nu}\,e^{\,2\,A(y)}\,.
\ee

\begin{figure}[ht!]
\vspace{1cm}
\includegraphics[{height=04cm,angle=00}]{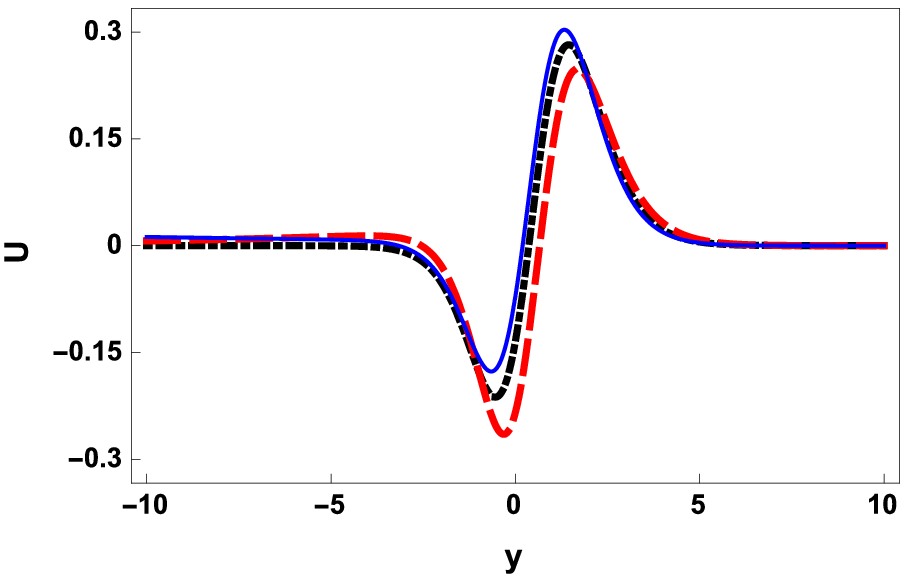} \hspace{0.1 cm} \includegraphics[{height=04cm,angle=00}]{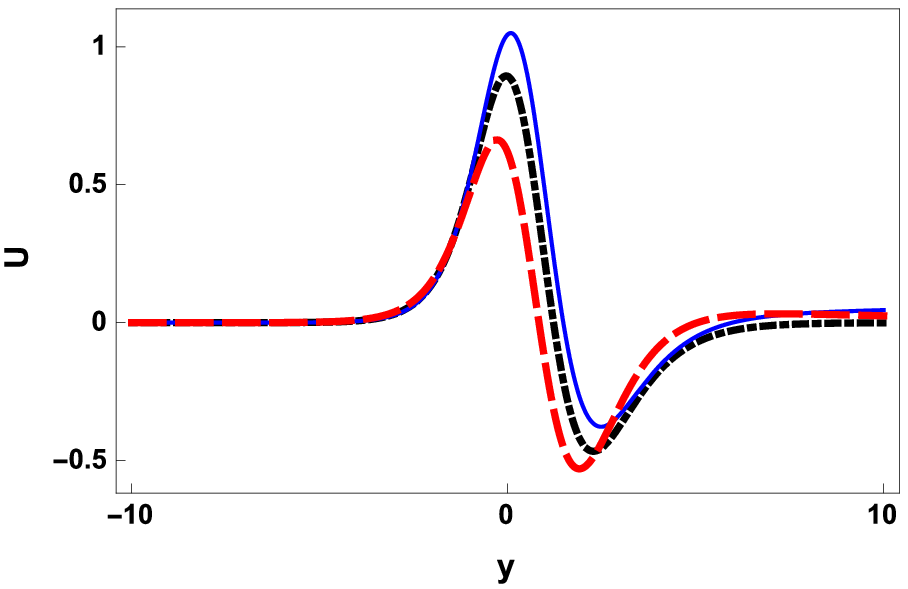}
\vspace{0.3cm}
\caption{In the left panel we see the effective potential, for $a=1/2$, $b=2$, $b_2=1$ (dotted black curve), $b_2=0.9$ (dashed red curve), and $b_2=1.05$ (solid blue curve). In the right we worked with  $a=1/2$, $b=2$, $b_2=-1/6$ (dotted black curve), $b_2=-1/6+0.1$ (dashed red curve), and $b_2=-1/6-0.05$ (solid blue curve).}
\label{FIG8}
\end{figure}

\begin{figure}[ht!]
\vspace{1cm}
\includegraphics[{height=04cm,angle=00}]{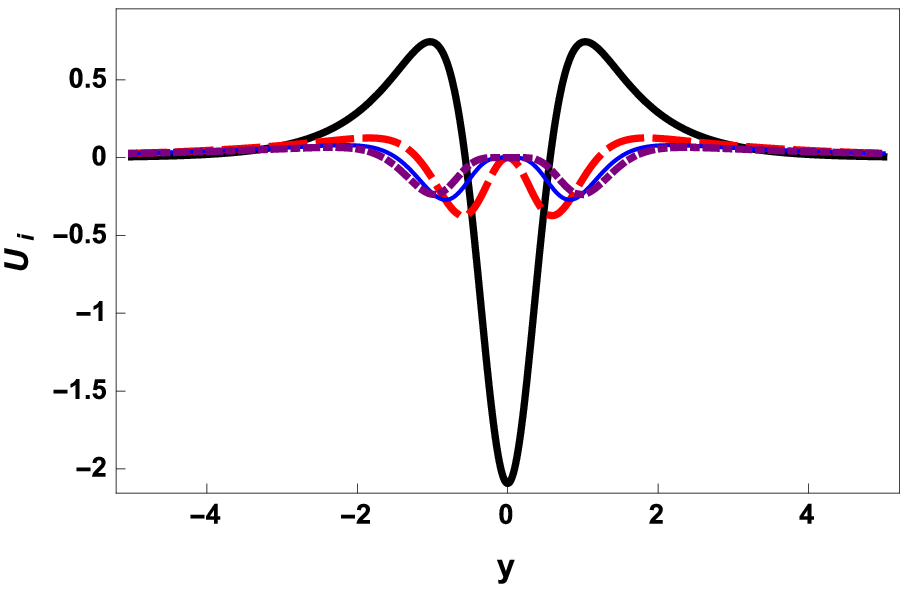} \hspace{0.1 cm} \includegraphics[{height=04cm,angle=00}]{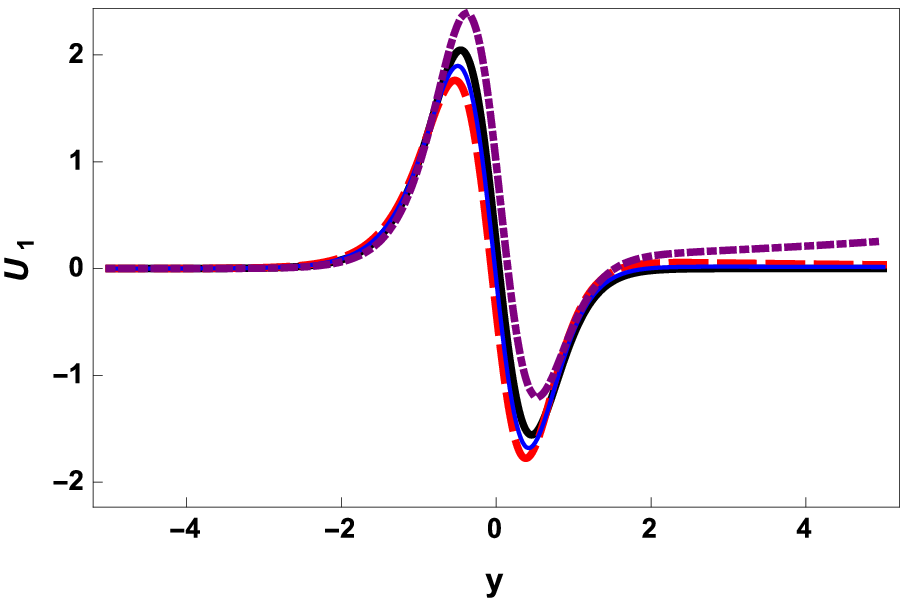}
\vspace{0.3cm}
\caption{In the left panel we see $U_i$, with $i=1$ (solid black curve), $i=2$  (dashed red curve), $i=3$ (solid thin blue curve) and $i=4$ (dotted dashed violet curve), for $a=1$, $b_2=-0.5$, and $\widetilde{c}=0$. The right frame shows $U_1$ with $a=1$, $b_2=-0.5$, $\widetilde{c}=2/3$ (solid black curve), $\widetilde{c}=2/3-0.1$ (dashed red curve), $\widetilde{c}=2/3-0.05$ (solid thin blue curve), and $\widetilde{c}=2/3+0.1$ (dotted dashed violet curve).}
\label{FIG9}
\end{figure}

\begin{figure}[ht!]
\vspace{1cm}
\includegraphics[{height=04cm,angle=00}]{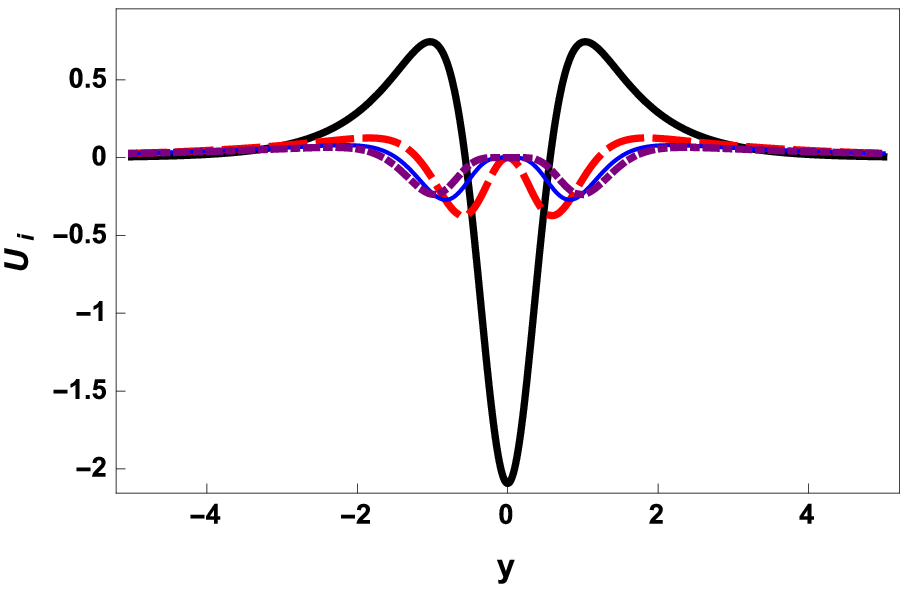} \hspace{0.1 cm} \includegraphics[{height=04cm,angle=00}]{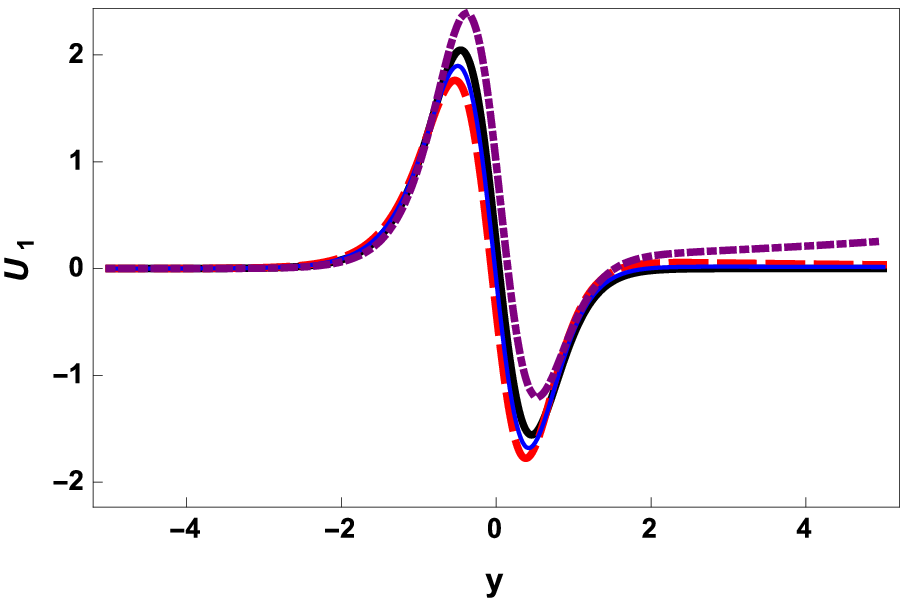}
\vspace{0.3cm}
\caption{In the left panel we see $U_i$, with $i=1$ (solid black curve), $i=2$  (dashed red curve), $i=3$ (solid thin blue curve) and $i=4$ (dotted dashed violet curve), for $a=1$,  and $\widetilde{c}=0$. The right frame shows $U_1$ with $a=1$, $\widetilde{c}=8/9$ (solid black curve), $\widetilde{c}=8/9-0.1$ (dashed red curve), $\widetilde{c}=8/9-0.05$ (solid thin blue curve), and $\widetilde{c}=8/9+0.1$ (dotted dashed violet curve).}
\label{FIG10}
\end{figure}

In Figs. \ref{FIG8}, \ref{FIG9} and \ref{FIG10} we can see the effective confining potentials for examples I, II, and III respectively, besides, some of them exhibit volcano shapes as shown in the left panels of Figs. \ref{FIG9}, and \ref{FIG10}. In Fig. \ref{FIG8}, the dotted black curve corresponds to the effective confining potential related with a critical brane, while the dashed red curve stands for the case of a non-critical brane, and the blue solid curve describes the supercritical brane regime. Moreover, in Figs. \ref{FIG9}, and \ref{FIG10} we present the solid black curve as the effective potential for a critical brane, while the dashed red curve and the solid thin blue curve correspond to non-critical branes, besides, the dotted-dashed violet curve stands for the supercritical brane. As the reader can see, the graphics from Fig. \ref{FIG9}, and \ref{FIG10} became more symmetric as we take $\widetilde{c}$ closer to zero. Those potentials which can localize gravity together with the zero-mode solutions, characterizes the stability of the branes (even for the asymmetric ones), except for the critical, and supercritical configurations, as expected. 

The stability for such brane configurations is supported by the study of fluctuations on a scalar gravity background, which was beautifully done in the seminal paper of DeWolfe et al. \cite{dewolfe/2000}. There it is possible to observe that the fluctuations due to the scalar field background decouple from the gravity fluctuations when the transverse traceless gauge is considered. Unless one considers generalizations of the Einstein-Hilbert action in five dimensions, which is going to change the equation of motion for the graviton, such a stability condition also works for hybrid braneworld models as one can see in the study of Veras et al. \cite{Veras/2016}. Besides, vectorial and scalar perturbations in a standard five-dimensional Einstein-Hilbert action coupled with $n$ scalar fields were investigated in details by Dutra et al. \cite{Dutra/2015}. There, the authors found that neither vectorial modes either scalar perturbations would affect the stability of such braneworlds.


\pagebreak

\section{Discussions}
\label{sec:discuss}

In this work, we verified how the extension method can be successfully applied to hybrid braneworld scenarios. The methodology based on superpotential function followed the steps of references \cite{bg,bls}. Such an application led us to construct more general two field braneworld models, besides it allowed us to recover some well-known results presented in \cite{bg}. Moreover, we could derive both symmetric and asymmetric branes for each one of the examples, beyond the critical case. These regimes were predicted by imposing the asymptotic behavior for the derivative of the warp-factor shown in \eqref{sec3_eq12_01}.  The characterization of the braneworld families here studied, presents non-trivial contributions from the coupling between the two scalar fields used in the extension method. Such non-trivial contributions emerge when we choose the parameter $b_2\neq 0$, and can be observed in the warp-factors, in the energy densities, in the Ricci scalars, and in the effective confining potentials for the graviton.

In the context of a standard Einstein-Hilbert action in five dimensions, the stability is normally associated with the existence of the zero-mode, which guarantees the existence of four-dimensional gravity localized on the brane \cite{dewolfe/2000}. The critical cases, however, do not enjoy the existence of normalizable zero modes, then, in principle, no four-dimensional gravity is localized on these branes. However, as shown in a series of papers \cite{Fonseca:2016yxw, Fonseca:2012bw, Fonseca:2011ep, Fonseca:2010va, Bazeia:2006ef}, the asymmetric branes with no zero modes can localize four-dimensional gravity through massive gravity modes. This is known as metastable gravity, firstly reported in \cite{Dvali:2000hr, Gregory:2000jc}. We hope to bring in near future stability analysis from the perspective of such metastable scenarios. Moreover, the presented methodology generalizes the results derived from Bazeia and Gomes in \cite{bg} and can be applied to other braneworld investigations presented in the literature. It would be interesting to study the cosmological bounds such hybrid models could impose, in the same spirit of the investigation done by \cite{Santos/2018}.  Another interesting issue would explore the extension method applicability in the context of modified five-dimensional gravity theories, such as Gauss-Bonnet model \cite{Gabriel/2014}, or even in different geometric scenarios for the branes, such as the one previously studied by Bernardini et al. \cite{Bernardini/2014} . We also believe that this procedure can be generalized to engender braneworlds composed of three or more fields, and we hope to report on some of these perspectives shortly.

\acknowledgments

We would like to thank CNPq, CAPES, and PRONEX/CNPq \& Paraiba State Research Foundation (Grants no.
165/2018, and 0015/2019), for partial financial support. FAB and JRLS acknowledge support from CNPq Grants no. 312104/2018-9, and 420479/2018-0, respectively. We also would like to thank the anonymous referees for their guidance and suggestions, which raised the quality of this work.


\begin{thebibliography}{99}

\bb{arkani_99} N. Arkani-Hamed, and M. Schmaltz, Phys. Rev. D {\bf 61}, 033005 (2000).

\bb{Gogberashvili/98} M. Gogberashvili, Int. J. Mod. Phys. D {\bf 11}, 1635 (2002).

\bb{Gogberashvili/99} M. Gogberashvili, Mod.Phys.Lett. A {\bf 14}, 2025 (1999).

\bb{rs_991}Lisa Randall, and R. Sundrum, Phys. Rev. Lett. {\bf 83}, 3370 (1999).

\bb{rs_992}Lisa Randall, and R. Sundrum, Phys. Rev. Lett. {\bf 83}, 4690 (1999).

\bb{thick_01} P. Kanti, I.I. Kogan, K.A. Olive, and M. Pospelov, Phys. Lett. B {\bf 468}, 31 (1999).

\bb{thick_02} Csaba Csaki, Joshua Erlich, Timothy J. Hollowood, and Yuri Shirman, Nucl. Phys. B {\bf 581}, 309 (2000).

\bb{thick_03} Aqeel Ahmed, and Bohdan Grzadkowski, JHEP {\bf 01}, 177 (2013).

\bb{thick_04} Aqeel Ahmed, Lukasz Dulny, and Bohdan Grzadkowski, The European Physical Journal C, {\bf 74}, 2862 (2014).

\bb{Bernardini/2014} Alex E. Bernardini, R. T. Cavalcanti, and Roldao da Rocha, Gen. Relat. Grav. {\bf 47},  1840 (2014).

\bb{thick_051} Nandinii Barbosa-Cendejas, Alfredo Herrera-Aguilar, Konstantinos Kanakoglou, Ulises Nucamendi, and Israel Quiros, General Relativity and Gravitation, {\bf 46}, 1631 (2014).

\bb{thick_052} Mariana Carrillo-Gonzalez, Gabriel German, Alfredo Herrera-Aguilar, and Dagoberto Malagon-Morejon, Gen. Rel. Grav. {\bf 46}, 1657 (2014).


\bb{Gabriel/2014} Gabriel German, Alfredo Herrera-Aguilar, Dagoberto Malagon-Morejon, Israel Quiros, and Roldao da Rocha, Phys. Rev. D {\bf 89}, 026004 (2014).

\bb{Dutra/2015} A. de Souza Dutra, G. P. de Brito, and J. M. Hoff da Silva, Phys. Rev. D {\bf 91}, 086016 (2015).

\bb{thick_07} D. Bazeia, and D. A. Ferreira, Annals of Physics {\bf 411}, 167975 (2019).

\bb{thick_08} D. Bazeia, D. A. Ferreira, and D. C. Moreira, EPL {\bf 129}, 11004 (2020).
 
\bb{bg} D. Bazeia, and A. R. Gomes, JHEP {\bf 0405}, 012 (2004).

\bb{dutra_13} A. de Souza Dutra, G. P. de Brito, and J. M. Hoff da Silva, Europhys. Lett. {\bf 108}, 11001 (2014). 

\bb{bmm} D. Bazeia, M. A. Marques, and R. Menezes, Phys. Rev. D {\bf 92}, 084058 (2015).

\bb{bls} D. Bazeia, L. Losano, and J.R.L. Santos, Physics Letters A {\bf 377}, 1615 (2013).

\bb{ms_14} P.H.R.S. Moraes, and J.R.L. Santos, Phys. Rev. D {\bf 89},   083516 (2014).

\bb{ms_18} J.R.L. Santos, P.H.R.S. Moraes, D.A. Ferreira, and D. C. Vilar Neta, Eur. Phys. J. C {\bf 78}, 169 (2018).

\bb{smb} J.R.L. Santos, D.S.S. Borges, and I.O. Moreira, EPL {\bf 123},  23001  (2018).

\bb{Dutra/2005} A. de Souza Dutra, Physics Letters B {\bf 626}, 249 (2005).

\bb{Chumbes/2009} Augusto E. R. Chumbes, and Marcelo B. Hott, 	Phys. Rev. D {\bf 81}, 045008 (2010).
 
\bb{Bazeia/2013} D. Bazeia et al., Phys. Scr. {\bf 87},  045101 (2013).

\bb{Almeida/2004} C. A. Almeida, D. Bazeia, L. Losano, and J. M. C. Malbouisson, Phys. Rev. D {\bf 69}, 067702 (2004).

\bb{bnrt}D. Bazeia, J. R. S. Nascimento, R. F. Ribeiro, and D. Toledo, J. Phys. A: Math. Gen. {\bf 30} 8157 (1997). 

\bb{blm} D. Bazeia, L. Losano, and J.M.C. Malbouisson, Phys. Review D {\bf 66},  101701 (2002).

\bb{bl2006} D. Bazeia, and L. Losano, Phys. Rev. D {\bf 73}, 025016 (2006).

\bb{almeida2014} W.T. Cruz, L.J.S. Sousa, and R.V. Maluf, and C.A.S. Almeida, Phys. Lett. B  {\bf 730} 314 (2014).

\bibitem{dewolfe/2000} O. DeWolfe, D.Z. Freedman, S.S. Gubser, and A. Karch, Phys.Rev.D {\bf 62}, 046008 (2000).

\bb{campos2002} A. Campos, Phys. Rev. Lett. {\bf 88}, 141602 (2002).

\bibitem{Veras/2016} D. F. S. Veras, W. T. Cruz, R.V. Maluf, and C. A. S. Almeida, Physics Letters B {\bf 754}, 201 (2016).


\bibitem{Fonseca:2016yxw} 
  R.~C.~Fonseca, F.~A.~Brito, and L.~Losano,
  arXiv:1611.03843 [hep-th].
\bibitem{Fonseca:2012bw} 
  R.~C.~Fonseca, F.~A.~Brito, and L.~Losano,
  Phys.\ Lett.\ B {\bf 728}, 443 (2014).
\bibitem{Fonseca:2011ep} 
  R.~C.~Fonseca, F.~A.~Brito, and L.~Losano,
  JCAP {\bf 1201}, 032 (2012).
\bibitem{Fonseca:2010va} 
  R.~C.~Fonseca, F.~A.~Brito, and L.~Losano,
  Phys.\ Lett.\ B {\bf 697}, 493 (2011).
\bibitem{Bazeia:2006ef} 
  D.~Bazeia, F.~A.~Brito, and L.~Losano,
  JHEP {\bf 0611}, 064 (2006).
  
\bibitem{Dvali:2000hr} 
  G.~R.~Dvali, G.~Gabadadze, and M.~Porrati,
  Phys.\ Lett.\ B {\bf 485}, 208 (2000).
\bibitem{Gregory:2000jc} 
  R.~Gregory, V.~A.~Rubakov, and S.~M.~Sibiryakov,
  Phys.\ Rev.\ Lett.\  {\bf 84}, 5928 (2000).
  \bibitem{Santos/2018} M. A. Santos et al., JCAP {\bf 03}, 023 (2018).





\end{thebibliography}
\end{document}